\documentclass[sigconf]{acmart}
\usepackage{booktabs}
\usepackage{tabularx}
\usepackage{xcolor,colortbl}
\definecolor{lightgray}{gray}{0.93}
\definecolor{slightgray}{gray}{0.98}
\definecolor{darkgray}{gray}{0.77}
\usepackage{makecell}
\usepackage{multirow}
\usepackage{enumitem} 

\newcolumntype{Y}{>{\centering\arraybackslash}X}
\AtBeginDocument{%
  \providecommand\BibTeX{{%
    \normalfont B\kern-0.5em{\scshape i\kern-0.25em b}\kern-0.8em\TeX}}}

\setcopyright{acmcopyright}
\copyrightyear{2024}
\acmYear{2024}
\setcopyright{acmlicensed}\acmConference[CHI '24]{Proceedings of the CHI Conference on Human Factors in Computing Systems}{May 11--16, 2024}{Honolulu, HI, USA}
\acmBooktitle{Proceedings of the CHI Conference on Human Factors in Computing Systems (CHI '24), May 11--16, 2024, Honolulu, HI, USA}
\acmDOI{10.1145/3613904.3642229}
\acmISBN{979-8-4007-0330-0/24/05}

\begin{document}

\title{ChatScratch: An AI-Augmented System Toward Autonomous Visual Programming Learning for Children Aged 6-12}
\renewcommand{\shorttitle}{ChatScratch}

\author{Liuqing Chen}
\authornote{Corresponding author.}

\affiliation{%
  \institution{Zhejiang University}
  \city{Hangzhou}
  \country{China}}
\email{chenlq@zju.edu.cn}

\author{Shuhong Xiao}
\affiliation{%
  \institution{Zhejiang University}
  \city{Hangzhou}
  \country{China}}
\email{xiao_sh@zju.edu.cn}

\author{Yunnong Chen}
\affiliation{%
  \institution{Zhejiang University}
  \city{Hangzhou}
  \country{China}}
\email{chen_yn@zju.edu.cn}

\author{Ruoyu Wu}
\affiliation{%
  \institution{Beijing Normal University}
  \city{Beijing}
  \country{China}}
\email{roylalala726@gmail.com}

\author{Yaxuan Song}
\affiliation{%
  \institution{Zhejiang University}
  \city{Hangzhou}
  \country{China}}
\email{songyx23@zju.edu.cn}

\author{Lingyun Sun}
\affiliation{%
  \institution{Zhejiang University}
  \city{Hangzhou}
  \country{China}}
\email{sunly@zju.edu.cn}







\renewcommand{\shortauthors}{Chen et al.}

\begin{abstract}
As Computational Thinking (CT) continues to permeate younger age groups in K-12 education, established CT platforms such as Scratch face challenges in catering to these younger learners, particularly those in the elementary school (ages 6-12). Through formative investigation with Scratch experts, we uncover three key obstacles to children's autonomous Scratch learning: artist's block in project planning, bounded creativity in asset creation, and inadequate coding guidance during implementation. To address these barriers, we introduce ChatScratch, an AI-augmented system to facilitate autonomous programming learning for young children. ChatScratch employs structured interactive storyboards and visual cues to overcome artist's block, integrates digital drawing and advanced image generation technologies to elevate creativity, and leverages Scratch-specialized Large Language Models (LLMs) for professional coding guidance. Our study shows that, compared to Scratch, ChatScratch efficiently fosters autonomous programming learning, and contributes to the creation of high-quality, personally meaningful Scratch projects for children. 
\end{abstract}




\maketitle

\section{Introduction}


\begin{figure*}[t]
    \centering
    \includegraphics[width=0.9\textwidth]{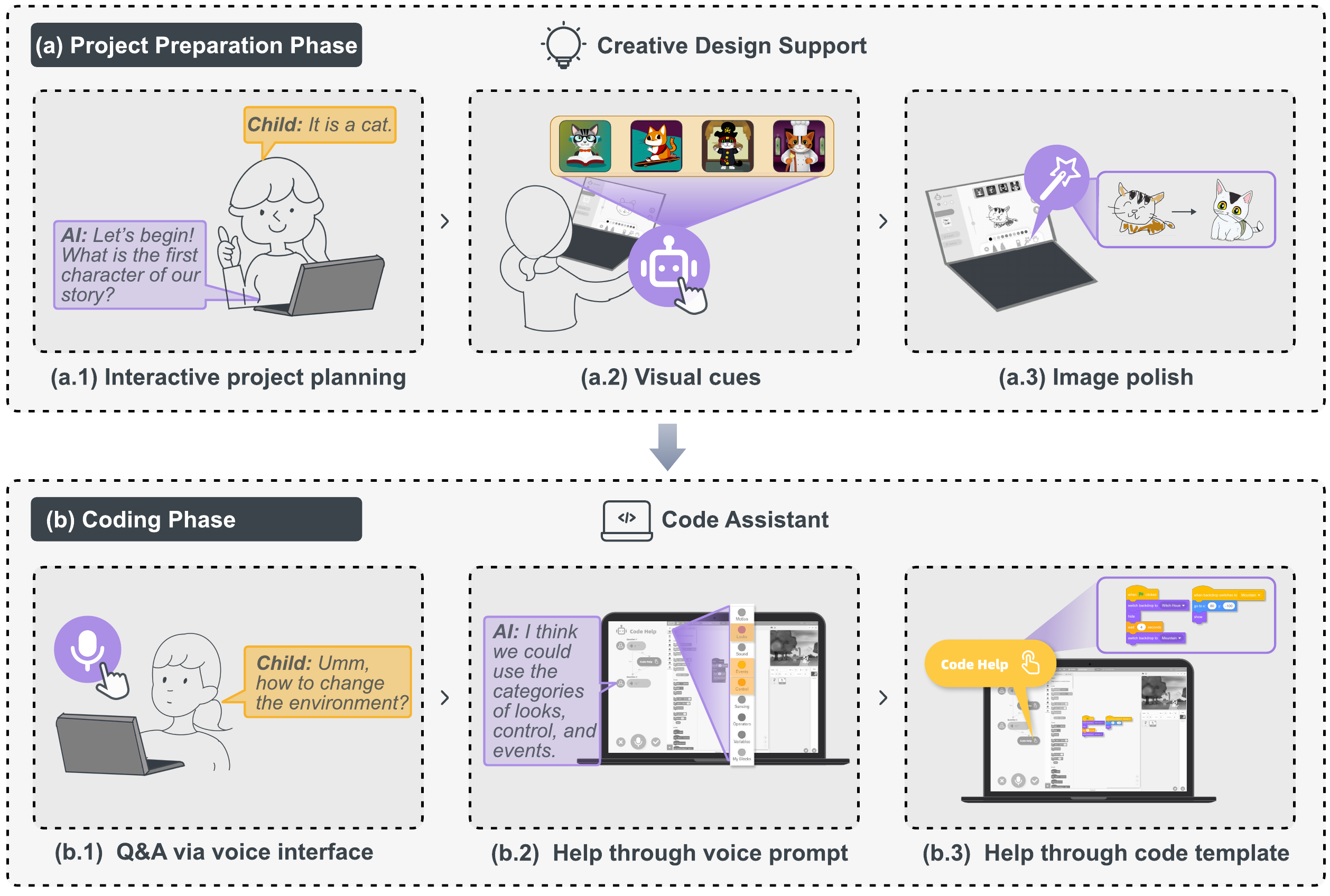}
    \caption{\textbf{Overview of ChatScratch: in the project preparation phase, ChatScratch assists children in planning their project with an interactive storyboard (a.1), provides visual cues to overcome their artist's block (a.2), and enhances asset quality with image polish (a.3). During the coding phase, it facilitates coding Q\&A via a voice interface (b.1), offers support through voice guide (b.2) and generates a foundational code template (b.3).}}
    \vspace{-0.25in}
    \label{fig:overview}
    \Description{The overview of ChatScratch, which is delineated in two main phases: the project preparation phase and the coding phase. During the project preparation phase, users employ an interactive storyboard provided by the system to craft the narrative of their project. With the assistance of a drawing board and an AI drawing helper, users can generate higher-quality and expectation-matching material images through a generative process. The system's built-in creative support aids users in their creative endeavors by supplying visual cues. In the coding phase, children interact with the system through voice-based Q&A to resolve implementation issues encountered during programming activities. The system offers verbal block code prompts or places a simple implementation plan in the code script area to accommodate users of varying abilities.}
    \vspace{0.15in}
\end{figure*}

In the past two decades, there has been significant traction in promoting Computational Thinking (CT) across all levels of schooling, from kindergarten through the $12^{th}$ grade (K–12). Scratch, due to its ability to provide teenagers with a more accessible and less abstract programming experience, has become the most popular CT educational tool, boasting over 100 million registered users from more than 150 countries \cite{ScratchStatistics}. However, as CT education shifts toward making it accessible to younger audiences—specifically, elementary school children between the ages of 6 and 12 \cite{grover2013computational}—platforms such as Scratch face new challenges. Young learners often lack the requisite literacy, arithmetic capabilities, fine-motor skills, and hand-eye coordination to effectively engage with these CT platforms \cite{SU2023100122}. Additionally, current Scratch educational practices are mainly tailored for classroom environments, guided by professional educators, and a structured curriculum \cite{10.1145/1226969.1227003}. Such an educational model is often inaccessible to students in under-resourced communities or developing countries. Although these students can, in theory, avail themselves of publicly available resources online, research shows that autonomous learning without the structured guidance of a classroom is often less effective \cite{10.1145/3544548.3580981,10.1145/1592761.1592779,10.1145/1999747.1999796}.


While early exposure to CT education has been recognized as beneficial for children's future career and life prospects \cite{lonka2018phenomenal}, it is crucial to create effective learning method for young learners, especially those without professional guidance or in home-based educational settings. One prominent approach in earlier research has been the adaptation of existing CT tools. For instance, ScratchJr \cite{papadakis2016developing} simplifies Scratch's user interface and reduces the number of code blocks, thereby lowering cognitive barriers for young learners. Additionally, Visual StoryCoder \cite{10.1145/3544548.3580981} eliminates hands-on coding activities, focusing solely on instilling an understanding of programming concepts such as loops, events, and variables. While there are numerous alternatives, Scratch remains the most commonly used platform in children's programming education, primarily because of its inherent tinkerable nature, profound meaningfulness, and vibrant social community \cite{10.1145/1592761.1592779}. Concurrently, another line of research, has bypassed tool-specific discussions to focus instead on the topic of learning strategies, where project-based learning emerges as the most promising approaches \cite{SU2023100122,hsu2018learn}. Personally meaningful projects, which allow children to freely choose their themes and forms of implementation, align exceptionally well with Scratch \cite{10.1145/1592761.1592779} in autonomous learning environments \cite{10.1145/3544548.3580981}. For the narrative consistency, we refer to the process of creating personally meaningful projects as theme-based creative programming in the following sections. 


Indeed, there are still barriers impeding children from engaging in theme-based creative programming with Scratch autonomously. To understand the challenges children encounter with Scratch, we conducted a formative investigation with six experienced Scratch educators. These educators, with an average of 6.67 years teaching experience, shared challenges faced by children and insights on how they navigate these challenges. In summary, we reported three main challenges: (1) children frequently get stuck suddenly, encountering an artist’s block—a temporary creative impasse where they struggle to progress with ideas \cite{10.1145/3334480.3382976} during the project planning stage; (2) the existing options for assets creation within Scratch constrain children's creative expression, leading to a form of bounded creativity \cite{romeiro2015bounded}; (3) the cognitive skills required for code implementation surpass the capabilities of children in their formal operational period, leaving them uncertain about where to start \cite{wertsch1984zone}. Inspired by these findings, our motivation is to provide the necessary support for children to cultivate CT skills through theme-based creative programming in Scratch, particularly in autonomous learning environments. 

Drawing on related work and our formative investigation, we introduce ChatScratch (Figure \ref{fig:overview}): an AI-augmented system towards autonomous programming learning for younger children. ChatScratch offers comprehensive support for the entire programming process. During the project preparation phase, ChatScratch leverages a structured interactive storyboard (a.1) with visual cues (a.2) as inspirations to overcome artist's block, enhancing both the detail and richness of the projects. When creating assets, ChatScratch incorporates Stable Diffusion with ControlNet \cite{zhang2023adding}, enables children to craft high-quality, tailored assets using image polish (a.3). As children embark on the coding phase, ChatScratch proposes a Scratch-specialized large language model to address their coding queries through voice interface (b.1). Through voice-guided code assistant (b.2), children are directed to find the required code blocks, enhancing their proficiency with code over time. By generating editable code templates (b.3) to the scripts area, it offers a tangible start point, guiding young learners to produce high quality code. 

We conducted a within-subject study with 24 children aged 6-12 to highlight the benefits of ChatScratch in fostering autonomous programming. Quantitatively, our study showed marked improvements in the creative aspects of children's programming, evidenced by an increase in the count of visual elements, a higher creativity support index and expert ratings. CT scores, along with measures of code retention and expansion, underscored ChatScratch's effectiveness in enhancing the programming process. Qualitatively, analysis of videos and interviews indicated that ChatScratch played a significant role in enabling children to create projects with personal meaning. In our discussion, we explored various benefits associated with ChatScratch using generative artificial intelligence (GAI), the seamless integration of creativity and coding in visual programming environments, and identified current limitations and potential future work.

Specifically, with ChatScratch, we present the following contributions:
\begin{itemize}

\item Through a systematic formative investigation with expert Scratch educators, we identified barriers hindering children's autonomous Scratch learning and reveal a heavy reliance on professional guidance. 

\item We presented ChatScratch, an AI-augmented system designed for autonomous programming learning among primary school children aged 6-12. Leveraging visual cues, image polish, and a code assistance, ChatScratch empowers children to craft personally meaningful projects, concurrently fostering their computational thinking skills.

\item We organized a comparative study between Scratch and ChatScratch on young children, employing both qualitative and quantitative methods to demonstrate the effectiveness of ChatScratch in enhancing children's autonomous Scratch learning experience.







\end{itemize}

\section{Related Work}
\subsection{Learning Computational Thinking through Programming}

Wing \cite{10.1145/1118178.1118215}, as the proponent, defined Computational Thinking (CT) as the thought processes that enable individuals to solve problems, design systems, and understand human behavior with principles fundamental to computer science. As Wing did not provide a precise definition for this term, the subsequent two decades witnessed intensive debates and discussions around CT \cite{selby2013computational}. Although a universally accepted core definition is still elusive, there is a consensus in the academic community that CT can be nurtured and refined through targeted training \cite{selby2013computational}. Programming, in particular, has been highlighted as an optimal medium for CT education. This claim is supported by two pieces of evidence: firstly, understanding the foundational principles and practices of programming has demonstrated benefits for an individual's academic and professional lives \cite{lonka2018phenomenal}. Secondly, the nuances of CT can be effectively evaluated through programming exercises \cite{10.1145/3105910}.

In both the realm of child education and the Human-Computer Interaction (HCI) domain, K-12 has become a prominent topic in discussions surrounding CT and programming \cite{10.1145/1118178.1118215,lonka2018phenomenal,10.1145/3544548.3580981}. Historically, educational practices and researches primarily focused on the high school level \cite{LYE201451}. However, in recent years, there has been an increasing emphasis and attention directed towards kindergarten and elementary school stages \cite{grover2013computational,10.1145/1118178.1118215}. Traditionally, the journey into programming commenced with mastering a programming language, such as Python and C. This approach, while rigorous, could often appear daunting and abstract to young learners \cite{ching2018developing}. To bridge this gap, new methodologies were introduced that featured graphical interface \cite{10.1145/1592761.1592779}, block-based programming with lenient syntax requirements \cite{flannery2013let}, and real-time execution \cite{weintrop2019block}. Building on these methodologies, tools such as LEGO \cite{klassner2003lego} and Scratch \cite{10.1145/1868358.1868363} emerged, enabling young learners to utilize programming to craft novel entities like interactive stories, games, and animations \cite{10.1145/1592761.1592779}. When leveraging these platforms to program, young learners are not simply learning to code; rather, they are coding to learn \cite{resnick2013learn}. 

The assessment of CT is crucial to understand not only the extent of children's individual skill development but also the broader impact of educational approaches such as Scratch \cite{ALLSOP201930}. There are multiple approaches to assessing CT, including project rubrics \cite{denner2014pair,moreno2015dr}, task-based questions \cite{duncan2015pilot,grover2015designing}, interviews \cite{mueller2017assessing}, and observations \cite{fessakis2013problem}. In this work, we employ Dr. Scratch \cite{moreno2015dr} for assessing children's CT performance in Scratch programming. Drawing from traditional programming assessment methods \cite{roman2019combining}, it aligns core CT concepts \cite{brennan2012new} with specific Scratch code blocks. As a representative project rubrics approach, it facilitates a quantifiable assessment by counting each kind of block, thereby providing an objective and consistent evaluation of the projects. Due to its intuitive interface and ease of accessibility, Dr. Scratch has emerged as the most popular assessment tool within the community. By leveraging its web service, learners can upload their Scratch projects and obtain instant score-based evaluations, eliminating the need for specialized expertise to understand their CT proficiency.

\subsection{Enhancing Creativity in Scratch Learning Environments} 
In discussing Scratch, it extends beyond the traditional view of programming as merely a coding skill. Instead, Scratch integrates coding into the everyday creative activities of children, incorporating storytelling, drawing, and gaming as fundamental forms of expression \cite{10.1145/1592761.1592779}. This underscores that Scratch is not solely a medium for fostering CT; it also necessitates a framework for supporting creative expression. Furthermore, the interplay between creativity and CT is noteworthy: the foundational elements of CT also serve as essential cognitive tools pivotal for creativity \cite{yadav2017fostering}. Thus, the nurturing of CT can be greatly enhanced in programming environments that facilitate creativity \cite{roque2016supporting}. 

Prior research has highlighted the significance of offering additional creative support within programming environments. For instance, Robertson et al. \cite{robertson2007adventure} implemented a creative process model in game design, which provides children with enhanced support and feedback during open-ended programming tasks. Howland et al. \cite{howland2015narrative} showcased Narrative Thread, a storytelling-based game design tool that streamlines the development of more complex game characters, dialogues, and events. Further studies, as indicated in \cite{10.1145/3544548.3580981, soleimani2019cyberplayce}, have also validated the role of storytelling in enhancing creative capacities when learning CT concepts. Leveraging the inherent structure of stories and their capacity to drive creative engagement \cite{land2013full}, we incorporate storytelling into ChatScratch as a foundational element to enhance children's project planning and asset creation. 

Expanding our perspective to a broader learning environments, the utilization of creativity support tools (CSTs) to aid in children's education and development has garnered widespread attention. Some studies have focused on the potential links between the design features of digital technologies and children's creative behaviors, offering valuable insights in the development of ChatScratch. For instance, Kucirkova et al. \cite{kucirkova2015child} observed that, when provided with computer-assisted drawing tools, children demonstrated more creative theme selections compared to traditional paper-based drawing methods. Inspired by this work, we have integrated computer-assisted drawing tools into ChatScratch, aiming to foster similar levels of creativity and exploration among young users. Echoing the idea of technology-enhanced creativity championed by Dietz et al. \cite{10.1145/3544548.3580981}, we also have explored methods to enhance tool support for children's personalized projects in ChatScratch, acknowledging that such tailored approaches heighten levels of creativity. Furthermore, we have adopted a similar scaffolding strategies used by Zhang et al. \cite{10.1145/3491102.3501914}, aimed at alleviating the burden on children during multiple tasks and enhancing their engagement.

\subsection{Supporting Children's Personally Meaningful Projects through Child-AI Collaboration}

In Scratch, personally meaningful projects refer to projects that reflect children's personal interests, emotions, and experiences. Normally, they are not restricted to a single paradigm and can manifest in various forms such as stories, animations, and games. Examples of these projects are abundant within the Scratch community \cite{scratch2023}. Yet, when it comes to real-world teaching, educators approach these projects with caution. The primary reasons being the deeper programming understanding they demand, which can be daunting for children, and also the additional effort and time teachers need to invest for individualized guidance and assessment. As a workaround, many educators lean towards standardized prescriptive projects \cite{delgaty2009curriculum}. These are often fragmented tasks, crafted to meet specific instructional objectives or curriculum requirements. While they ensure uniform learning benchmarks, they often neglect to foster students' exploration and identification of personal interests. Moreover, for children without access to educational resources or guidance from educators, learning within paradigms without classroom support is often ineffective \cite{10.1145/3544548.3580981,10.1145/1592761.1592779,10.1145/1999747.1999796}. Recognizing the challenges, and understanding the undeniable value of personally meaningful projects, the potential of Child-AI Collaboration systems presents a viable alternative. These systems aim to bridge existing gaps with tailored guidance, while preserving the core of personal exploration and learning.

Utilizing AI as a companion for children's learning and entertainment, aiding in fostering positive behaviors, has already established a mature foundation within the realm of Child–Computer Interaction (CCI). Generally, we can observe two distinct forms from prior works: tangible entities and virtual agents. Tangible entities, exemplified by PopBots \cite{williams2018popbots}, Amazon's Alexa \cite{AmazonAlexaDeveloper}, and RoBoHoN \cite{robohon_official_website}, afford the benefit of physical interactivity, allowing children a tactile, hands-on learning experience. In contrast, virtual agents, such as Bio Sketchbook \cite{10.1145/3459990.3465197}, StoryCoder \cite{10.1145/3544548.3580981}, and Teachable Machine \cite{TeachableMachine} offer greater flexibility. They can be incorporated into electronic devices like tablets, thus broadening the scope of application scenarios. Furthermore, they permit seamless integration into targeted educational frameworks, such as the strategies employed by StoryDrawer \cite{10.1145/3491102.3501914} to support children in visual storytelling without altering the original workflow. In addition to the discussion on tangible and virtual platforms, voice interfaces stand out as a distinctive feature in Child-AI collaboration. They not only assist children in maintaining attention \cite{10.1145/3311927.3323151} but also dismantle literacy barriers by eliminating the need for written text \cite{10.1145/3544548.3580981}.


To support children's autonomous learning, ChatScratch needs to offer a wide range of expertise, spanning areas such as programming and storytelling. Furthermore, strong adaptive capabilities and affective qualities are also considered essential for the efficacy of AI assistants \cite{lin2020parental}. Despite these ideals, previous AI assistants designed for children have often fallen short. Some systems \cite{10.1145/3544548.3581043} rely on digital sensors to perceive real world information but are limited to rudimentary interactions, and other approaches have relied on predefined rules to generate dialogue, which often led to logical inconsistencies \cite{zhang2022storybuddy}. Large language models (LLMs), however, present a compelling alternative. The extensive corpus of data equips them with remarkable capabilities in natural language understanding and fluid expression \cite{OpenAIChat2023}, thereby substantially enhancing the interactive performance of AI assistants. Moreover, they are conducive to specialized training for domain-specific knowledge and tasks, providing targeted and accurate guidance. \cite{touvron2023llama}. These qualities elevate LLMs to a powerful assistant in promoting children's autonomous learning.

\section{Formative Investigation}
In this section, we present our formative investigation with six experienced Scratch educators. Our objectives were two-fold: (1) to identify the challenges children face when using Scratch for autonomous programming learning through theme-based creative programming and (2) to discover design opportunities for ChatScratch that address these challenges. We first outline the process through which the formative investigation was conducted. We then present our key findings and implications for design. Finally, we summarize the design objectives for our ChatScratch.


\subsection{Method}
We recruited six expert Scratch educators (3 females, 3 males) of age 26-39 (M=31, SD=4.44) with Scratch teaching experience range from 4 to 12 years (M=6.67, SD=2.96). All the educators were affiliated with a tutoring company in China that specializes in programming for children. Our decision to engage with professional educators instead of children and parents was driven by the understanding that educators still hold an irreplaceable role in programming education \cite{10.1145/1226969.1227003}. Their expertise in consolidating various educational materials, tools, and practices is pivotal for effectively nurturing CT skills in young children \cite{https://doi.org/10.1111/bjet.12355}.

First, we attended some educators' offline programming classes in the tutoring center, where each classroom could accommodate 6-15 students. Here, the center provided computers for each student, facilitating their programming activities. The educators introduced us as guest visitors to the children, and ensure our presence did not alter the natural flow of the classroom. During these classroom sessions, we observed the children as they tackled theme-based creative programming tasks. We documented the challenges they faced, their responses, and overall performance. Furthermore, we observed the educators' strategies, noting how they provided guidance and addressed issues as they arose. Each session lasted about 120 minutes, including a brief 10-minute break in between. After each session, three researchers convened to discuss the notes, identifying key topics for subsequent interviews with the educators.  

In the following two weeks, we scheduled one-on-one semi-structured interviews with each educator. One of the interviews was conducted face-to-face, while the others were carried out via online meetings, depending on the participants' preferences. The interviews lasted 41-54 min (M=47.83, SD=4.18). Guided by our previous classroom observations, we structured the interviews around the workflow of children's theme-based creative programming, specifically on three key phases: project planning, assets creation, and code implementation. Moreover, we also discussed with educators about the format of theme-based creative programming and their perspectives on how AI can assist in children's programming. All interview sessions were conducted in Chinese and were recorded with the consent of the participants. After the interview, each participant received a gift valued of \$20 as compensation. 

We analyzed the interview transcripts with a reflexive thematic analysis \cite{braun2006using,braun2021one}. All interview recordings were transcribed into text using Whisper, and then translated into English by GPT-4. Subsequently, a researcher manually segmented the materials based on the three key phases discussed. For each segmented data, we coded the challenges and possible solutions introduced by educators. Finally, for each phase, we established several themes that encapsulate both the perspectives of the educators and our insights. For both interviews questions and the results of the thematic analysis, please refer to Appendix A in our supplementary materials.


\subsection{Findings and Design Implications}
In this section, we briefly present our findings, encompassing the themes created at each key phase, and the significance of supporting personally meaningful projects through theme-based creative programming.

\subsubsection{Artist's Block in Project Planning}
Traditional programming often begins with a well-defined task and a linear thought process, such as calculating the Fibonacci sequence in Python. In contrast, Scratch projects require a phase of creative thinking and conceptualization before one starts coding \cite{10.1145/1592761.1592779}. While children naturally tend to either bypass this planning stage or integrate it directly into the coding process, all six educators underscore its importance, directing children through this critical step to help crystallize their project. During this preparatory step, children clarify vague ideas and formulate a project plan. For instance, if a child envisions crafting a fairy tale within Scratch, their would first establish the foundation by basic concepts such as characters and scenes. Subsequently, they would embellish the narrative with additional details to vividly portray the story. Yet, there is a hurdle: most young children are at the formal operational stage of cognitive development, which often limits their abstract reasoning and executive functions, essential for constructing a cohesive project plan \cite{barrouillet2015theories}.


This challenge was evident when one of our interviewees recounted assisting an 8-year-old girl with her Scratch project, which was about \textit{``an experience of watching a waterfall''}. Having described the main character and the waterfall, the girl found herself at an impasse and hit what is commonly known as an artist's block \cite{10.1145/3334480.3382976}. She had aspirations of adding more characters and expanding the story's setting. However, Scratch offered minimal imaginative support, hindering her ability to discover the appropriate details to enhance the narrative. Furthermore, while she had a profound emotional impression of the waterfall's beauty, she faced challenges in translating this abstract emotional experience into a detailed description of waterfall element for her Scratch project. To past this hurdle, four interviewees mentioned employing a simple storyboard to conceptualize their stories and seeking inspiration to stimulate creativity.



Building upon these observations, our design approach centered around two pivotal insights. First, we recognized the importance of providing a structured narrative framework to help children clarify the foundational elements of their projects \cite{10.1145/3450741.3465254}. Second, it is crucial to provide stimuli that can ignite creative thinking in children, aiding them in the detailing and refinement of their stories; such stimuli can help children overcome artist's block \cite{10.1145/3544548.3580948,10.1145/3411763.3450391}. By providing this support, we provide them with a clear reference point, thereby enabling them to systematically build upon their initial ideas and ensuring a cohesive development of their projects.

\subsubsection{Bounded Creativity in Assets Creation}
The second major challenge children face when using Scratch often centers around the design and creation of assets. Although in the actual implementation this process is usually interwoven with project planning, we discuss it separately here due to its unique challenges and significant impact. As a visual programming platform, Scratch heavily relies on a diverse array of assets to help users manifest their creative visions. Among these, sprites and backdrops stand out as the most prevalent, corresponding to the characters and environments within a child's narrative framework. 

Scratch offers various methods for asset creation. First, children can import materials externally, utilizing resources from the internet or remixing from other projects. While this may seem like a versatile feature, in practice, \textit{``it is less frequently employed by younger users''}, as reported by one interviewee. The process of sourcing, selecting, adapting, and importing external assets can be daunting and complicated, particularly for those who are still developing their digital literacy skills \cite{10.1145/3459990.3460696}. Beyond the technical aspects, this detour into searching and handling process can serve as a cognitive interruption \cite{10.1145/1314683.1314689}. When a child is deeply engaged in their project, diverting attention to search for and process assets can interrupt their creative momentum, potentially leading to disjointed ideas or diminished enthusiasm for the project. The second and more prevalent method is leveraging Scratch's built-in assets library. However, Scratch's built-in assets do not always align with children's creative visions. An interviewee highlighted a situation where a child had imagined a unique character but could not find a match in the library. Consequently, they settled for the default Scratch Cat. This scenario illustrates the concept of bounded creativity \cite{romeiro2015bounded}, where predefined resources might inadvertently limit a child's imaginative expression. As this interviewee noted:

\begin{quote}
`\textit{`I have seen many kids use the Scratch Cat as their character, not by choice, but due to the lack of appropriate assets. Such limitations frequently lead to their frustration as their original creative intent is not truly represented.'' }
\end{quote}

In fact, we found children's prioritization of assets exceeded our initial expectations. Through an interviewee, we learned that many children consider good Scratch projects to be primarily characterized by delicate assets. Some children view coding as a means to animate their assets, with the desirable assets having the potential to inspire their enthusiasm for coding. In practice, educators sometimes encourage children to create assets through drawing. The advantage lies in the fact that through drawing, children can more effectively achieve their ideas and personality. However, weaknesses are also evident. Drawing requires more classroom time, and quality assurance is not guaranteed, as reported by one interviewee. 

Following these observations, a pivotal design inspiration emerged for ChatScratch: it should facilitate children's unrestricted expression during asset creation, ensuring their original visions are realized without compromise. Specifically, we chose drawing as the medium for assets creation, recognizing its intuitive and straightforward nature for children in comparison to other digital literacy \cite{10.1145/3491102.3501914, 10.1145/3544548.3580981}. Moreover, considering the limitations in children's drawing abilities and the aspiration for high-quality, tailored assets, we incorporated Stable Diffusion with ControlNet \cite{rombach2022high,zhang2023adding}. Using this approach, children can produce assets simply from brief descriptions and simple doodles, thus effectively translating their creative visions into reality.

\subsubsection{Code Assistance as a Scaffold}

The coding process in Scratch is similar as playing with LEGO bricks: each code snippet works as a unique building block that can be pieced together to construct more complicated functions \cite{10.1145/1868358.1868363}. While Scratch uses text-based descriptions to make each code block more understandable for children \cite{10.1145/1227504.1227388}, its coding system remains complex for young learners. For instance, selecting a specific block from the vast array of over a hundred options, categorized into eight groups, is still a significant challenge. As mentioned by one interviewee, even children with over a year of Scratch experience frequently struggle to find their targeted block. 

Furthermore, five interviewees unanimously reported a dilemma among their students: these fresh programmers frequently feel daunted at the beginning. They anticipate guidance from their teachers to get started; otherwise, they often engage in protracted deliberation. Typically, educators provide verbal instructions or hands-on demonstrations for foundational implementation. Such a foundational code guidance provides not only a tangible starting point but also instills a greater sense of confidence for children. Intriguingly, when equipped with such a foundation, children often produce outstanding projects that impress educators. In fact, a foundational code solution can be viewed as a classic embodiment of the scaffolding strategy \cite{10.1145/3544548.3580935}. Drawing from Vygotsky's Zone of Proximal Development (ZPD), children's coding learning is optimized as it occurs within the range of tasks they can perform with support of such a baseline but cannot master independently \cite{wertsch1984zone}. It offers them the necessary support to overcome initial hesitations about where to start and encourages more proactive exploration and expansion of their coding endeavors. 



With these findings, new insights emerged for our ChatScratch. Our system should provide young learners with access to basic coding solution as a starting point, helping them overcome the initial feeling of being overwhelmed and also enables learners to expand upon and refine their programming ideas. To tackle the challenge of locating specific code blocks and to make the interaction more engaging, we decided to honor the children's imaginative request, as conveyed by one of our interviewers: ``\textit{it would be great if the code could just appear in the script area}.'' By doing this, we aim to assist them in effortlessly generating simple, executable code templates as basic solution.

\subsubsection{Support Personally Meaningful Projects}
Unlike standardized projects that focus primarily on imparting specific knowledge or programming concepts, personally meaningful projects are more likely to reflect the child's individual interests, experiences, or aspirations.

As presented by all six interviewees, children consistently show a stronger affinity for projects that hold personal meaning for them. The reasons are twofold. First, personally meaningful projects significantly enhance children's engagement, thereby encouraging creative behavior, deepening their understanding, and amplifying their intrinsic motivation to explore and innovate \cite{reid2007design}. Second, children exhibit a strong sense of pride and accomplishment when they engage with personal meaningful projects. According to one interviewee, children are more inclined to discuss and share projects that encompass their own life stories, hobbies, or fantasies. They resonate more deeply with the sentiment, ``\textit{This is something I created}'', fostering a stronger sense of ownership and achievement \cite{10.1145/3544548.3580981}.  

So far, the design elements we have introduced for assisting children in project planning, asset creation, and code development have essentially laid the foundation of personally meaningful projects. Beyond these foundational elements, our interviewees have also imparted some overarching design principles and interaction considerations that can further enrich our system design. First, they underscored the significance of preserving the original Scratch workflows to avoid introducing additional cognitive burdens. This implies that our code scaffolding should be integrated as a separate, parallel procedure. Second, our interviewees also emphasized the imperative nature of a transparent and structured workflow for facilitating autonomous learning. In response to this, we have integrated an intuitive navigational interface and provide tutorial video to elucidate the operational flow of our system.

\subsection{Design Goals}
Building upon previous research and insights from our investigations, we derived three principal objectives for our system: 

\begin{itemize}
\item Goal 1 aims to guide children during the project planning phase, providing structured storyboards and visual cues to ensure they have a rich and detailed project plan.
\item Goal 2 centers on supporting children in the asset creation phase, equipping them with the means to produce tailored assets with high quality.
\item Goal 3 focuses on guiding children through code development offering high-quality code template for a smooth start. 
\end{itemize}

Overall, we aim for ChatScratch to address the prevalent challenges children face in autonomous Scratch learning, guiding them to improve CT skills by crafting personally meaningful projects.

\section{The design of ChatScratch}
\begin{figure*}[htp]
    \centering
    \includegraphics[width=0.9\textwidth]{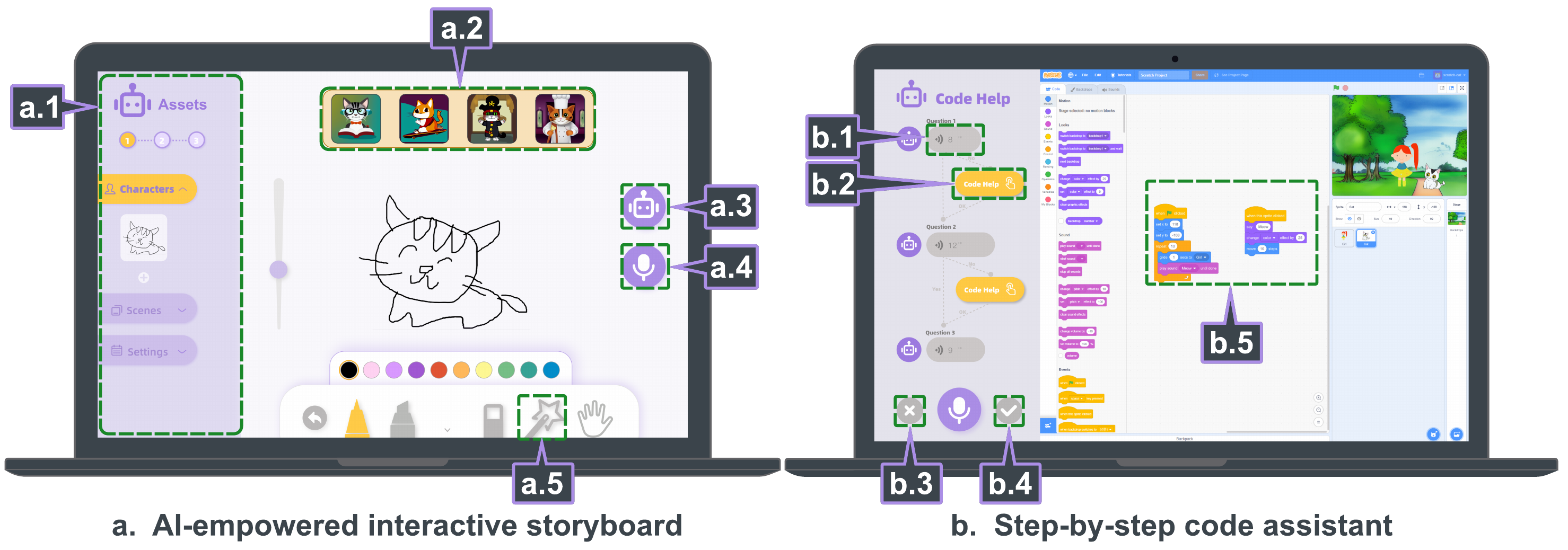}
    \caption{\textbf{When using \textit{ChatScratch}, the child first (a) plans and creates assets with an AI-empowered interactive storyboard (i.e., (a.1) storyline, (a.2) visual cues, (a.3) inspiration button, (a.4) voice button, (a.5) image polish button), then (b) programs with step-by-step code assistant (i.e., (b.1) dialog bubbles button, (b.2) ``code help'' button, (b.3) ``unclear'' button, (b.4) ``clear'' button, (b.5) the generated Scratch blocks). }}
    \vspace{-0.25in}
    \label{fig:UI}
    \Description{Two image labelled (a) and (b) depict the user interfaces of the ChatScratch system. In image (a), the interactive storyboard for project preparation is shown, which includes the (a.1) storyline panel on the left, (a.2) visual cues located at the top of the drawing board on the right, (a.3) inspiration button, and (a.4) voice button situated next to the drawing board, along with the (a.5) image polish button found within the toolbar below the drawing board. In image (b), the coding interface is illustrated, with the (b.1) dialog bubbles button located in the dialogue bar on the left, which assists in acquiring solutions. This panel also includes the (b.2) "code help" button for coding assistance, the (b.3) "unclear" button for signaling when further help is needed, and the (b.4) "clear" button for terminating the current task. The (b.5) generated Scratch blocks appear in the script area to the right.}
    \vspace{0.15in}
\end{figure*}

\subsection{System Overview}

Based on these three design goals, we built ChatScratch (Figure \ref{fig:UI}). Powered by advanced image generation technologies and large language models (LLMs), ChatScratch assists children in project planning, asset creation, and programming, thereby facilitating autonomous learning of Scratch for young children. Our ChatScratch consists of two user interfaces including an interactive storyboard and a step-by-step code assistant. Figure \ref{fig:UI}.a displays the interactive storyboard interface primarily aimed at facilitating project planning and asset creation for children. The step-by-step code assistant (Figure \ref{fig:UI}.b), maintains the original Scratch GUI while incorporating a Scratch-specialized large language model (LLM). Leveraging a Scratch-specialized LLM, the assistant can not only assist children by generating step-by-step code guidance but also improve their CT skills and engagement.

\subsection{Goal 1: Interactive Structured Storyboard and Visual Cues}

Following the typical classroom practice where experienced educators often provide children with grid paper to draft storyboards for project planning, we created an explicit project planning framework using a storyline (Figure \ref{fig:UI}.(a.1)) for our system. Leveraging principles of task decomposition and abstraction thinking \cite{shute2017demystifying}, the storyline is designed to guide children in deconstructing the story into three basic acts. Each act consists of components including characters, scenes, and events. When creating these story components, children can press the voice button (Figure \ref{fig:UI}.(a.4)) to verbally relay their descriptions to the voice agent and use the drawing board to simply sketch out their current story components. 

As we discovered through our interviews, young children often encounter cognitive obstacles during the planning phase, facing what we term as artist's block \cite{10.1145/3334480.3382976}. To address this challenge, the interactive storyboard (Figure \ref{fig:pipeline_1}) employed visual cues as stimuli to help children detail and enrich the project. First, children can click the inspiration button (Figure \ref{fig:UI}.(a.3)) to detail their project. For instance, they might say, \textit{``My character is a cat.''} In such cases, children can click on the inspiration button (Figure \ref{fig:UI}.(a.3)), and the system will display four refined concept images on the canvas: \textit{a cat reading with glasses}, \textit{a skateboarding cat in casual wear}, \textit{a cat in a pirate hat on a ship}, and \textit{a cat in chef attire}. Second, we address the need for children to enrich their projects. In many cases, children may already have some initial settings for their projects but find it challenging to add new characters or scenes that make the project more diverse and engaging. For example, they might say, \textit{``I have no idea about the next character.''} Children can click on these visual cues to hear the explanations of the voice agent, including a basic description of the new element and how it relates to the existing elements in the project.

It's worth noting that in both scenarios, we do not mandate that children should follow our recommendations. On the contrary, these visual cues allow children the freedom to observe, combine, and transfer these ideas according to their creative instincts. This design not only helps children articulate their ideas more clearly but also fosters the development of their abstract thinking and attention to details.

\subsection{Goal 2: Drawing-based Assets Creation with Advanced Image Generation}

As revealed in our formative investigation, children often find it challenging to acquire assets that match their ideas, whether through external imports or built-in libraries. In ChatScratch, we introduced a drawing board to facilitate asset creation through digital drawing. Children can draw and erase on the canvas, select colors, and adjust line widths using the drawing tools. Considering the limitations in children's drawing skills and the need for high-quality assets, we implemented advanced image generation technologies based on the stable diffusion model \cite{StableDiffusion2023} to polish their doodles.

\begin{figure*}[htp]
    \centering
    \includegraphics[width=0.9\textwidth]{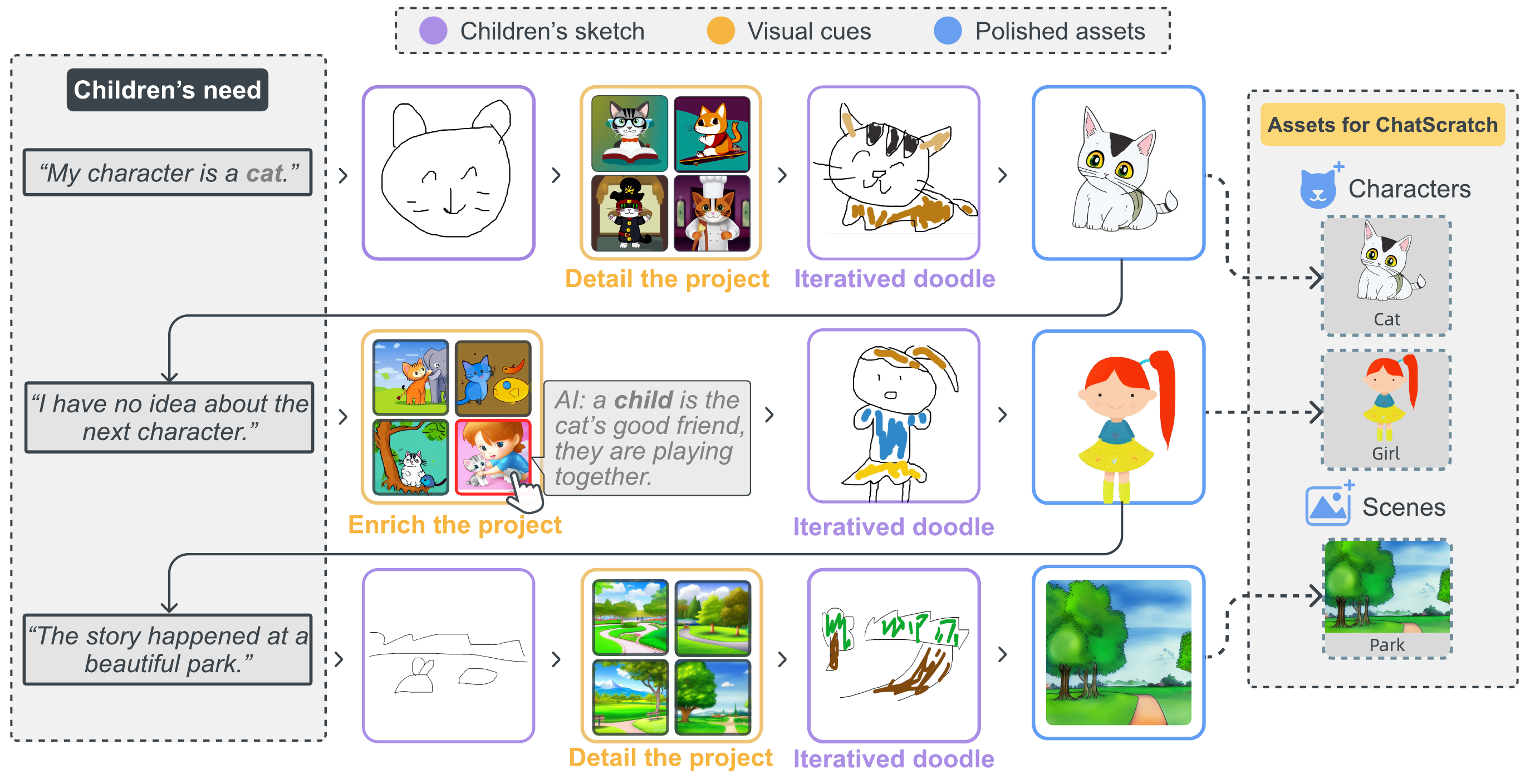}
    \caption{\textbf{Overview of the assets creation pipeline for ChatScratch. Children express their needs and sketch on the drawing board. Children can then obtain visual cues to detail and enrich their assets. The stable diffusion model and ControlNet were used to iteratively polish their doodles. All polished assets are automatically imported into the ChatScratch programming interface.}}
    \vspace{-0.15in}
    \label{fig:pipeline_1}
    \Description{The creative support in asset creation. Based on the user's simple doodles and verbal descriptions, the system generates image objects that align with the requirements to serve as drawing references. Users can refine details such as line and color based on these references and use the image polish button to produce high-quality assets. The process is iterative, allowing users to seamlessly switch between doodling and asset generation, making modifications as needed. The system also fosters creative inspiration for new asset creation. Drawing from the user-defined assets, it generates associated material images, providing specific relational details. This feature enables users to expand their project with contextually relevant and visually cohesive elements.}
    \vspace{0.1in}
\end{figure*}

To strike a balance between high-quality output and preserving a personal creative touch, we employ ControlNet \cite{zhang2023adding} to extract key points from children's doodles, typically capturing essential colors, outlines and positional features. In addition, we utilize large language models (LLMs) to distill characters and scenes from children's descriptions, converting them into prompts that are compatible with our image generation model. Figure \ref{fig:pipeline_1} showcases examples of polished assets, wherein the essential outlines and positional information provided by the child are well-preserved. The image prompts generated by LLMs served to enhance the quality of the polished asset. During the asset creation process, children have the option to repeatedly activate the polishing feature by clicking the ``image polish'' button (Figure \ref{fig:UI}.(a.5)), allowing for iterative refinement until they are satisfied with the asset. This approach provides children with greater control over the final output, thereby further reinforcing their sense of ownership and engagement with the project. After completing the asset creation task, all materials are automatically imported into the coding interface, eliminating the manual efforts typically required by children in Scratch.

We also pay close attention to the quality and suitability issues during the assets generation process. The Frechet Inception Distance (FID) score \cite{chong2020effectively} and Sliced Wasserstein Distance (SWD) \cite{kolouri2019generalized} score are utilized as technical evaluations to ensure accurate and natural asset creation for children. FID is crucial for gauging the similarity in style and content between two image sets, where a lower FID score signifies greater resemblance. Similarly, SWD measures the textural and structural parallels between image sets, with a lower score indicating better similarity. We collected materials from both the built-in library and the Scratch community, and calculated the FID and SWD scores between them as references. Comparatively, the FID and SWD scores between the built-in library assets and those generated by our ChatScratch demonstrated superior outcomes. This indicates that the quality of the assets generated by ChatScratch is higher than those found in the community. For detailed experimental procedures and results, please refer to supplementary materials appendix B. Additionally, we also make efforts to ensure the generated content is also suitable and safe for children. For the LLM used to generate image prompts, we implemented a role-playing strategy \cite{Shanahan2023Role}, explicitly informing it to assume the roles of an educator and a child assistant. This approach is intended to tailor its outputs to be educational and child-friendly. Moreover, we integrated negative prompts \cite{Kapoor2023NegativePrompts,Stone2023NegativeAIPrompts} into the generation process, explicitly excluding keywords such as violence, horror, sex, crime, and discrimination. This methodology aims to further safeguard the generated content, ensuring it is suitable and safe for young audience.

\subsection{Goal 3: Code Assistant supported by Scratch-specialized Large Language Model}

The code assistant is designed to understand the coding confusion of children, analyze what type of code blocks children need, and guide children on how to get started with Scratch programming. Once children had navigated through the storyline, finished their project planning, and prepared their digital assets (such as characters and scenes), the code assistant would say: \textit{``Excellent job! Now, it's time to click on the coding button. Let's start on our coding journey!''} If children cannot convert inspiration into specific code logic, they can express the programming problems to the code assistant. If the child gets stuck, the code assistant will encourage the child to use it, by saying \textit{``Press the voice button and tell me about your problem, and together we will come up with a solution.''}

\begin{figure*}[t]
    \centering
    \includegraphics[width=0.95\textwidth]{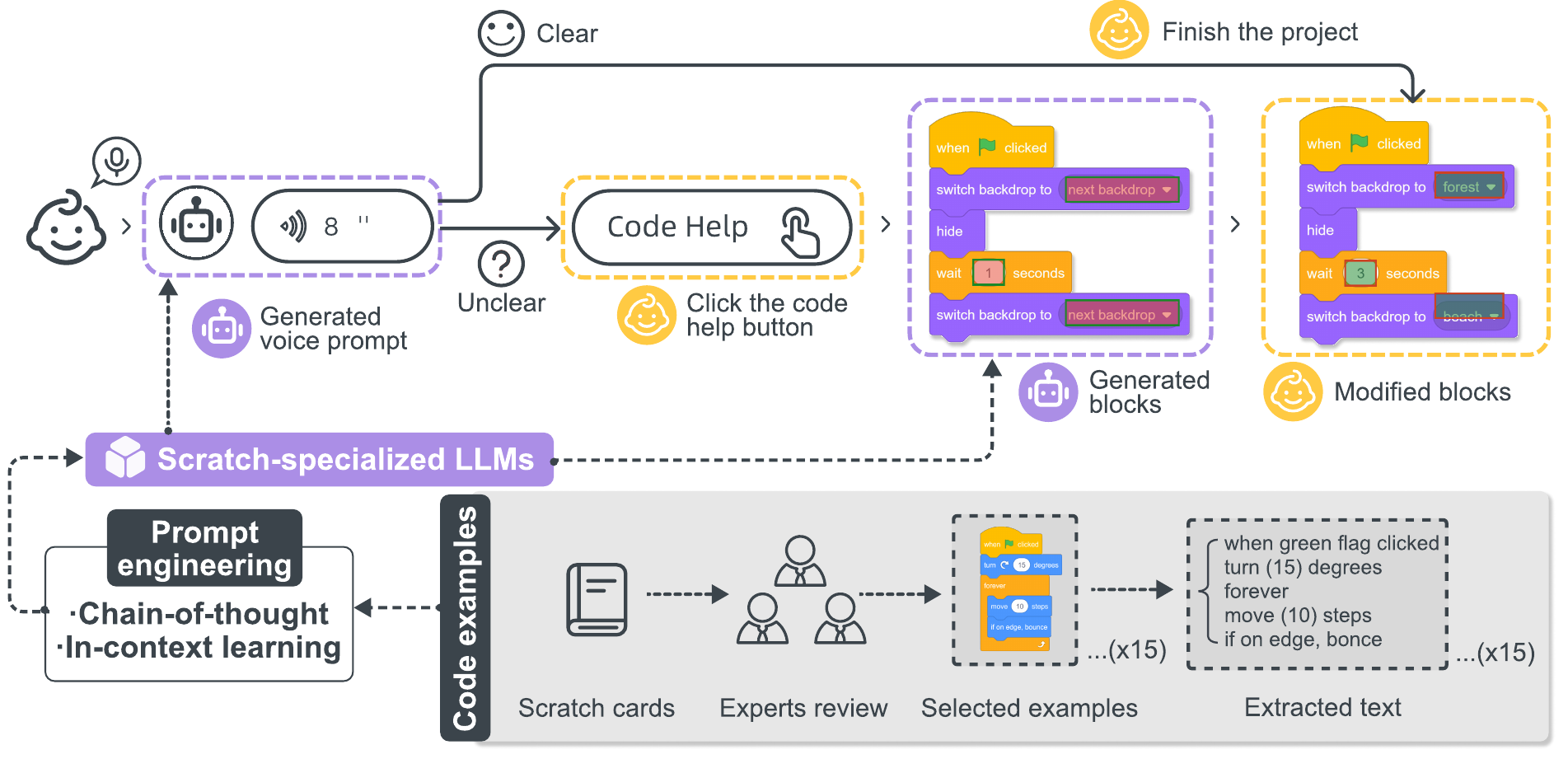}
    \caption{\textbf{Overview of the code assisting pipeline for ChatScratch. Children can ask the code assistant which is supported with a Scratch-specialized large language model, to get step-by-step coding tips. In the first step, the assistant generates voice prompts on how to use related Scratch blocks. In the second step, the assistant generates Scratch block templates to provide a tangible starting point. To generate high-quality code templates, we use expert-selected samples and employ prompt engineering to train our large language models.}}
    \vspace{-0.15in}
    \label{fig:code_pipeline}
    \Description{The overview of the code assisting pipeline for ChatScratch. The bottom half of the image depicts the creation process of the Scratch-specialized Large Language Model (LLM), which includes a structured process of Scratch knowledge based on expert tagging and knowledge input through prompt engineering, as mentioned in the main text. The top half of the image shows the user actions and corresponding model feedback: in response to children's queries, the Scratch-specialized LLM provides voice-based block code usage tips. If the user indicates understanding, the current dialogue task is concluded. If not, a "code help" button is presented, and a complete coding solution is generated.}
    \vspace{0.15in}
\end{figure*}

\begin{quote}
    \textit{CHILD: [Press the voice button] How to make the SpongeBob (created by children) jump?}

    \textit{AGENT: [Highlighting the categories in the Scratch block area] I think we can use the categories of motion, control, and event blocks}.
\end{quote}

Considering children's literacy level, such conversations will be presented as voice chats. As shown in Figure \ref{fig:code_pipeline}, we present the code assisting pipeline for ChatScratch. In the first step, children can click on the appearing dialog bubbles (Figure \ref{fig:UI}.(b.1)) to obtain suggestions for code block categories and how to use them. For beginners, they may still not be able to find the appropriate code blocks with only voice prompts. If so, in the second step, they can click the ``unclear'' button (Figure \ref{fig:UI}(b.3)) to tell the system that they need more specific help. Then children can click the appearing ``code help'' button (Figure \ref{fig:UI}.(b2)), the system will automatically generate a block combination that consists of multiple blocks to the Scratch programming area (Figure \ref{fig:UI}(b.5)) to give children a tangible starting point for programming. Children can press the ``clear'' button (Figure \ref{fig:UI}.(b.4)) at any time after asking a question to end the current dialogue.

Specifically, we proposed a LLM specialized for Scratch, which learns knowledge from a large-scale training corpus and is capable of recognizing and applying computational thinking concepts in classic Scratch programming cases. As discussed in \cite{rodriguez2020computational}, a visual programming practice can be expressed as the understanding of a single block and the effects that block combinations can produce. Therefore, we provided LLMs with a few examples of code blocks as in-context learning and a chain-of-thought as reasoning, which is illustrated in the following subsections. An example of the prompt is shown in Table~\ref{tab:prompt2}. With this in-context learning, we prompt the LLMs to generate block combinations in the same pattern based on children's questions.

\renewcommand{\arraystretch}{1.1}
\begin{table*}
\footnotesize
\centering
\caption{\textbf{The examples of prompt engineering of code generation.}}
\vspace{-0.04in}
 
	\label{tab:prompt2}
	\begin{tabular}{>{\bfseries} p{120pt}|p{360pt}} 
		\hline
		\rowcolor{darkgray} \bf{PROMPT TYPE}  & \bf{INSTANTIATION} \\
		\hline
            \hline
            \multicolumn{2}{l}{\textbf{Example 1}} \\
            \midrule
		Child's question & \textit{How to realize click on the rabbit and make it run all the time?} \\
		  \midrule
    	Chain-of-Thought & \textit{ 
          Solve a question-answering task with interleaving Thought. First, please select the answer from the Scratch 3.0 categories below Motion, Looks, Sound, Events, Control, Sensing, Operators, and Variables. Second, please provide your answer based on the Scratch Wiki Blocks.} \\
		\midrule
		The input example in LLM & \textit{Here are some examples. Question: How do I make a character walk from sitting to right? Answer: First, Use Motion to control character movement, Looks to switch character actions, and Control to repeat execution. Second, [ "when green flag clicked", "switch costume to [sitting]", "wait [2] seconds", "change x by (10)", "switch costume to [walking]"]. Question: [child's question]} \\
		\midrule

        The output example in LLM & \textit{First, use Events to trigger the action upon click, and Motion to control the movement of the rabbit. Additionally, Control to repeat the running action indefinitely. Second, a simple code snippet for this would be : ["when sprite clicked", "forever", "move [10] steps"]}  \\
            \midrule
            \multicolumn{2}{l}{\textbf{Example 2}} \\
            \midrule
		Child's question & \textit{How to make the box disappear when it hits the car?} \\
		  \midrule
    	Chain-of-Thought & \textit{{Solve a question-answering task with interleaving Thought. First, please select the answer from the Scratch 3.0 categories below Motion, Looks, Sound, Events, Control, Sensing, Operators, and Variables. Second, please provide your answer based on the Scratch Wiki Blocks.}} \\
		\midrule
		The input example in LLM & \textit{Here are some examples. Question: How to press space to make the character jump? Answer: First, use Events to detect the space key press event, Motion to control the character's movement for jumping, and Control to perform the sequence of actions that constitute a jump. Second, here's a simple code snippet for this: ["when space key pressed", "change y by [10]", "wait [0.5] seconds", "change y by [-10]"] Question: [child's question]} \\
		\midrule

        The output example in LLM & \textit{First, use Sensing to detect the collision between the box and the car, and Looks to control the visibility of the box. Second, here is an example code snippet for this: ["when green flag clicked", "forever", "if "touching [car]" then", "hide"]}  \\
        \bottomrule
	\end{tabular}

\end{table*}

\subsubsection{In-context Learning}

A representative example helps LLM elicit specific knowledge and abstractions necessary to accomplish the task \cite{yao2022react}. In the Scratch community, Scratch cards \footnote{\url{https://resources.scratch.mit.edu/www/cards/en/scratch-cards-all.pdf}} stand out as quintessential introductory materials for programming. The compactness and portability of Scratch cards render them an optimal resource for both newcomers and classroom courses. To select high-quality examples from Scratch cards, three educators with more than four years of Scratch teaching experience were hired to work with the research team. Each educator was asked to independently review all Scratch cards and selected samples with educational significance. After an initial review of one hour, each educator completely browsed through the Scratch cards, and on average selected 7 representative code examples. Subsequently, the three educators and two researchers convened to discuss and refine the data examples until consensus was achieved. Statistically, we have collected 15 code examples as our representative dataset. A code example for few-shot learning is shown at the bottom of Figure \ref{fig:code_pipeline}.

\subsubsection{Chain-of-thought Reasoning}

Though in-context learning proves its efficacy for basic tasks involving a couple of exemplars, it faces hurdles with complex tasks that demand logical reasoning and multi-step resolutions, such as arithmetic or common-sense reasoning problems \cite{wang2022rationale, kojima2022large}. To leverage the reasoning capabilities of LLMs, researchers have introduced the chain-of-thought (CoT) reasoning approach \cite{wei2022chain}, where rationales are integrated as intermediate steps with examples, as depicted in Table \ref{tab:prompt2}. To employ this paradigm, we asked the educators to independently write the code guidance with the chain-of-thought reasoning for each code example in the representative dataset. Then we discussed with educators how to map the text generated by LLMs onto specific Scratch blocks to achieve visual code generation. 

Table \ref{tab:prompt2} shows an example of chain-of-thought reasoning for prompting LLM. In the first step, we explicitly provided the LLM with code block categories. Subsequently, we prompted it to generate structured codes in a text format. In the first step, we prompted LLM by explicitly inputting the meaning of the categories of code blocks. In the second step, we prompted LLM to generate structured codes in text format. To match the text-based LLM responses with the actual code block IDs, we computed the Levenshtein distance between them. For example, if the model outputs the ``when sprite clicked'' text, the corresponding Scratch block can be matched according to semantics. This can prevent illusions caused by LLMs and improve the quality of the generated Scratch blocks. Besides, the coding assistant can foster children's CT skills, as they still need to continue modifying the generated blocks of code. The generated Scratch blocks (Figure \ref{fig:code_pipeline}) are just a rough starting point. Children still need to set the number of loops, change the delay time, and program interaction strategies.

\subsection{Implementation}
Overall, ChatScratch comprises a front-end web page built in React \footnote{\url{https://react.dev/}} and a back-end server developed in Python. To avoid literacy barriers, we utilized the Tencent Cloud Text-to-Speech API \footnote{\url{https://cloud.tencent.com/product/tts}} for voice agent to broadcast the voice prompts and the OpenAI speech recognition API \footnote{\url{https://openai.com/research/whisper}} for transcribing speech. In the front-end, we implemented our code assistant based on the open-source Scratch GUI \footnote{\url{https://github.com/scratchfoundation/scratch-gui}}. To achieve interaction between two user interfaces running independently on different ports of the server (i.e., interactive storyboard and Scratch GUI), we adopted HTTP redirection technology for page jumps. The back-end of ChatScratch was implemented with the Nginx web server and the Flask library \footnote{\url{https://github.com/pallets/flask}}. It receives output text from the LLMs and parses it with the Langchain parser \footnote{\url{https://www.langchain.com/}}. For LLMs, we use the state-of-the-art GPT4 model. In accordance with a small-scale pilot study, and to accommodate the inherent maximum input tokens of the LLMs (specifically 8196 for GPT4), we have designated the number of few-shot learning examples to be ten. To simulate the click and drag operations of Scratch GUI, we adopted webpage automation technology. By mapping the text output from LLMs to the code block element IDs in Scratch GUI, we are able to precisely select and operate the code block corresponding to the ID, thus realizing the visual code block generation process.
\section{Evaluation}

To assess the effectiveness of ChatScratch, we carried out a within-subjects study, comparing it with the original Scratch platform. Twenty-four children, aged 6-12, participated in our experiment. Each child experienced both systems, undertaking two theme-based creative programming tasks in separate sessions. Primarily, we aimed to address three research questions through our experiments: 

\begin{itemize}

\item RQ1: How does ChatScratch support creative tasks in Scratch programming?
\item RQ2: How does ChatScratch support high-quality code implementation in Scratch programming?
\item RQ3: How does ChatScratch support children to create personally meaningful projects autonomously? 

\end{itemize}

\subsection{Participants}
We recruited 24 children (10 female, 14 male) as participants aged 6-12 years (M = 8.96, SD = 2.12). Each child participated in two theme-based creative programming sessions, with a 48-hour interval between sessions. All participants were Scratch beginners and had less than 1 year of experience with Scratch. The participants were recruited via social media, and the experiment was approved by the university's ethics committee.All participants were native Mandarin speakers. As compensation for participating in the experiment, each child received a gift valued of \$20.

\subsection{Procedure}

\begin{figure}[htp]
    \centering
    \includegraphics[width=0.45\textwidth]{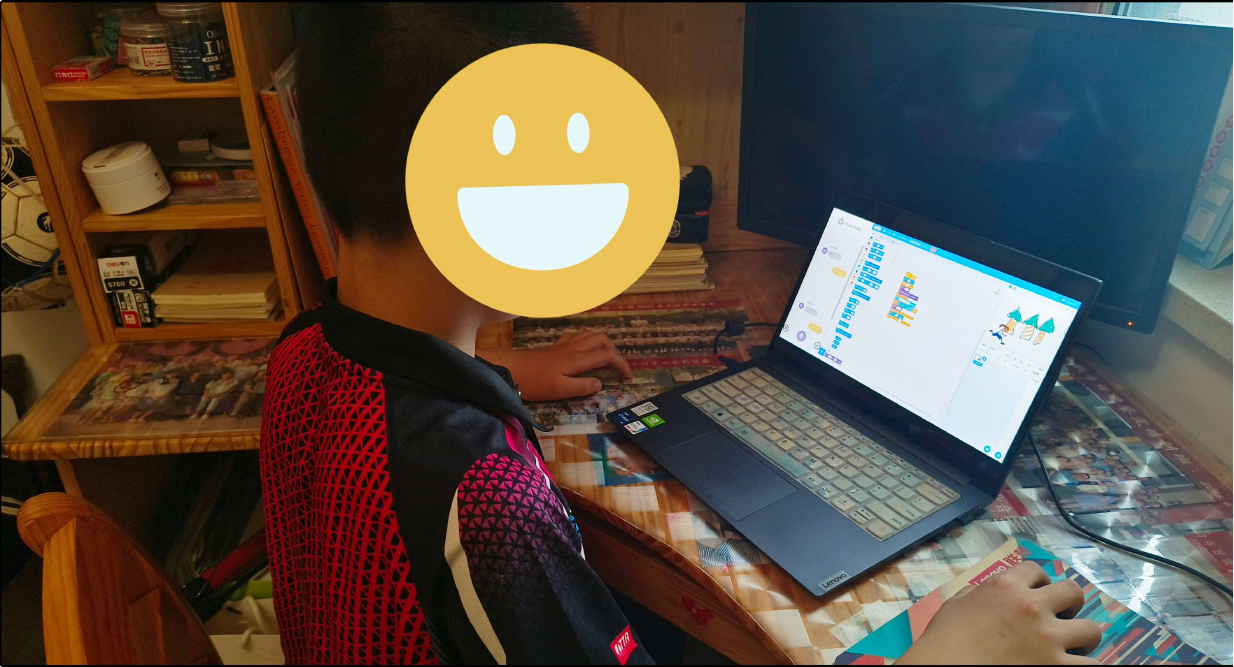}
    \caption{\textbf{Child participants' behaviors recordings. A boy was interacting with ChatScratch.}}
    \vspace{-0.11in}
    \label{fig:Recordings}
    \Description{A child participant engaged with ChatScratch on a laptop.}
\end{figure}

As a within-subjects study, each participant was required to undertake two theme-based creative programming sessions (Figure \ref{fig:Recordings}). To avoid potential time-related effects and carryover effects, we introduced a 48-hour interval between sessions. Additionally, we incorporated a counterbalanced design, alternating between combinations of tools (ChatScratch or Scratch) and themes (A or B) to ensure unbiased results \cite{vonDavier2004}. Inspired by previous studies \cite{10.1145/3491102.3501914,8534789} and suggestions from educators, we covered two specific themes: \textit{A: Memorable Experiences} and \textit{B: Animal Stories}. Each session spanned around 70 minutes—comprising 20 minutes dedicated to understanding system usage (with our tutorial videos for ChatScratch and built-in tutorial videos for Scratch) and free exploration, 10 minutes for participants to pose questions and clarify any doubts with the researchers, 10 minutes break, and a concentrated 30 minutes for the autonomous theme-based creative programming task. Both ChatScratch and Scratch were configured with a Chinese interface based on participant preferences. During the tasks, researchers were only allowed to assist children when unexpected technical difficulties arose. Following each session, we gathered video recordings from two sources: screen recordings that captured the software interactions and side-angle footage that documented participants' behaviors. In addition, we gathered the project files created by the children in the sb3 format. Subsequently, participants were asked to complete the Creativity Support Index questionnaire \cite{cherry2014quantifying} to gauge how effectively their processes were supported. After completing both sessions, we conducted a brief semi-structured interview with participants to delve into their user experience with ChatScratch for the theme-based creative programming tasks. 

\begin{table*}[htp]
\centering
\footnotesize
\caption{\textbf{Summary of the collected data and the evaluation metrics utilized.}}
\vspace{-0.11in}
\label{Collected Data and Metrics Summary}
\begin{tabularx}{0.95\textwidth}{p{0.23\textwidth}p{0.23\textwidth}X}
\toprule
Data & Evaluation Metrics & Description \\
\midrule
Visual Element Count& Assets Richness  & Count of sprites and backdrops in projects to quantify project richness \cite{10.1145/3491102.3501914}.\\
\addlinespace
Expert Ratings& Expert Ratings on Assets & Evaluation by Scratch educators to determine the originality, consistency, creativity, and quality of children's assets \cite{amabile1982social}. \\
\addlinespace
Creativity Support Index Questionnaires& Creativity Support Index & Gauges system's usability and effectiveness in enhancing creative tasks \cite{cherry2014quantifying}. \\
\addlinespace
Code Quality Rubric Scores& Dr. Scratch Rubric& Assess seven CT within children's Scratch code snippets \cite{moreno2015dr}.  \\
\addlinespace
Code Retention and Expansion& Retention and Expansion Measures& Evaluate how children use and build upon provided code templates.\\
\addlinespace
Video Recordings&Coding by Researchers & Examination of children's actions, identifying pauses, mistakes, and unintentional behaviors during system use. \\
\addlinespace
Artifact-based Interview& Semi-Structured Interview & Gather insights on children's project creation, their process, and feedback on ChatScratch \cite{portelance2015code}. \\
\bottomrule
\end{tabularx}
\end{table*}

\subsection{Collected Data and Metrics}
In this section, we describe the collected data and metrics leveraged for our evaluations. Collectively, we amassed data across seven distinct categories as summarized in Table \ref{Collected Data and Metrics Summary}. Out of these, five types—visual element count, generative quality of assets, expert ratings, Creativity Support Index questionnaires, code quality rubric score, and code retention and expansion—were earmarked for quantitative analyses. The remaining two, artifact-based interview and video recordings, were dedicated to qualitative insights. Any data requiring subjective evaluations were independently assessed by multiple evaluators, with consistency among them documented. 

\subsubsection{Visual Element Count}

Inspired by \cite{10.1145/3491102.3501914}, we adopted a visual element count metric to assess the richness of children's projects. Within our theme-based creative programming tasks, the visual element count is quantified as the sum of sprites and backdrops, reflecting the richness of a child's project planning.

\subsubsection{Expert Ratings}

The expert ratings were employed to evaluate both the depth of children's project planning and the performance of assets created during the asset creation phase. Our primary concern was to understand if, with the aid of ChatScratch, children could bring their concepts to fruition more freely, rather than being confined to limited creativity and resource support. To undertake this assessment, we followed Amabile's Consensual Creativity Assessment Technique (CAT) \cite{amabile1982social} and invited three Scratch educators to independently evaluate the children's creative outputs. Employing a 5-point Likert scale, the experts assessed the assets based on four primary criteria: (1) Originality: indicating the extent to which the assets reflects the child's own creation rather than being sourced from existing resources; (2) Consistency: indicating the extent to which the assets align with the child's descriptions during project planning; (3) Creativity: indicating the degree to which the assets demonstrate innovation and imaginative expression; (4) Quality: reflecting the inherent attributes of the materials, including completeness, clarity, detail, and harmony. The evaluations showcased a significant agreement among the experts, with an overall high interclass correlation coefficient (ICC) of 0.83 (p< .001).

\subsubsection{Creativity Support Index Questionnaires}
We employed the Creativity Support Index \cite{cherry2014quantifying} to gauge children's perception of ChatScratch's usability and its ability to facilitate creative tasks. This questionnaire comprises 12 questions, spread over six themes. Responses were captured using a 5-point Likert scale, allowing us to gather firsthand feedback from the children and insights into their experiences.

\subsubsection{Code Quality Rubric Scores}
We utilized the Dr. Scratch rubric \cite{moreno2015dr} to evaluate the code quality within children's theme-based creative programming projects. As a widely applied metric in evaluating Scratch code, it appraises code quality by quantifying seven CT dimensions:  abstraction, parallelism, logic, synchronization, flow control, interactivity, and data representation. The score for each dimension, ranging from 0 to 3, is determined by the count of corresponding code snippets. The scoring details can be found in our supplementary material. In essence, a heightened code quality rubric score suggests that children are more extensively harnessing and learning CT skills during their project development \cite{10.1145/3544548.3580981}. 

\subsubsection{Code Retention and Expansion}
To assess the effectiveness of our coding assistance, we devised the metrics of code retention and expansion. The retention rate \( R \) quantifies the degree to which children adopt the provided code, indicating its usability and the children's acceptance. On the other hand, the expansion rate \( E \) illustrates how children modify and build upon the code template, emphasizing how the initial code serves as an inspiration for children's coding. Let \( X \) denote the set of code snippets from our templates, and \( Y \) represent the set of code snippets in a child's final project. The measures are defined as:

\begin{equation*} \label{eq:formulas}
\begin{aligned}
    R &= \frac{|X \cap Y|}{|X|}, \quad
    E &= \frac{|Y| - |X \cap Y|}{|Y|}.
\end{aligned}
\end{equation*}

\subsubsection{Video Recordings}

For our study, we gathered two distinct types of video recordings: one that captured the on-screen operations and another that recorded the child's physical behaviors during the session. Through screen recordings, we focus on how children use the core features of ChatScratch, including the frequency of use, patterns, and fluency. For the recordings capturing the child's behaviors, we particularly focus on the discordant parts of the programming process, such as prominent pauses and errors. Video results and interview results were combined for a reflexive thematic analysis, focusing on how children's interactions differ when using ChatScratch as opposed to Scratch. For detailed results, please refer to Appendix D in our supplementary materials.


\subsubsection{Artifact-based Interview}
The artifact-based interviewing technique offers us a granular insight into whether our system effectively supports children in crafting personally meaningful projects. Drawing on prior work \cite{portelance2015code}, we structured our interview around three topics to understand the children's created projects (question 1-2), their project processes (question 3-4), and their perspectives on ChatScratch (question 5-7):
\begin{enumerate}
    \item Tell me about your projects.
    \item How did you come up with the idea? Was your idea realized?
    \item We noticed that during [specific time/event], you exhibited [specific behavior]. What problems did you encounter at that time?
    \item How did you address these challenges?
    \item How was your experience using ChatScratch?
    \item What aspects did you find impressive? Were there any parts that you found challenging?
    \item Would you be willing to use ChatScratch instead of Scratch?

\end{enumerate}

\subsection{Data Analysis}

\subsubsection{quantitative metric}
Despite implementing a counterbalanced design in our experiment to mitigate the effects of order and theme on the outcomes, we still verified through data analysis whether these factors were effectively controlled, thereby enhancing the rigor of our study. To achieve this objective, we treated all quantitative metrics as dependent variables, employing a MANOVA that incorporates factors including tool, theme, and order. We expect that significant differences in outcomes will only emerge from varying tools. Following this, paired t-tests are utilized to delve deeper into the distinctions between Scratch and ChatScratch. To mitigate the risk of Type I errors associated with multiple comparisons, a Bonferroni correction is applied. All the data and analysis codes are included, for details, please see Appendix E in our supplementary materials.


\subsubsection{qualitative metric}
Video recordings and interview results are utilized for a reflexive thematic analysis. In RQ3, we report on two themes related to creating personally meaningful projects. Other observations and supporting materials are presented in supplementary material D. These supplementary insights, although valuable, stem from and go beyond the primary exploration of this paper, thereby would serve as a source of inspiration for future research.



\section{Results}
In this section, we systematically answer the three research questions based on evidence from our data. First, we answer the question of how does ChatScratch support creative tasks in RQ1. Second, we explore how ChatScratch supports high-quality code implementation based on code assistance in RQ2. Lastly, in RQ3, we discuss how ChatScratch supports children in autonomously creating projects that hold personal meanings. As our MANOVA demonstrates, there is a significant effect on tool \(F(4,41) = 21.59, p < 0.01^{**}\), while no significance on order \(F(4,41) = 0.46, p = 0.76\) and theme \(F(4,41) = 0.19, p = 0.94\). This confirms that both order and theme were effectively controlled, allowing us to directly report the subsequent paired t-tests in RQ1 and RQ2.


\subsection{RQ1: How does ChatScratch Support Creative Tasks in Scratch Programming}

We employed the metric of visual element count to evaluate the efficacy of structured planning and visual cues in ChatScratch. The average count in ChatScratch was \(4.81\) (\(SD = 2.34\)), compared to \(3.06\) (\(SD = 1.34\)) in Scratch. A paired-samples \(t\)-test revealed a statistically significant difference between the two platforms in enriching children's projects, with \( t(23) = 5.40, p < 0.01^{**} \). 

\begin{figure*}[thp]
    \centering
    \includegraphics[width=0.8\textwidth]{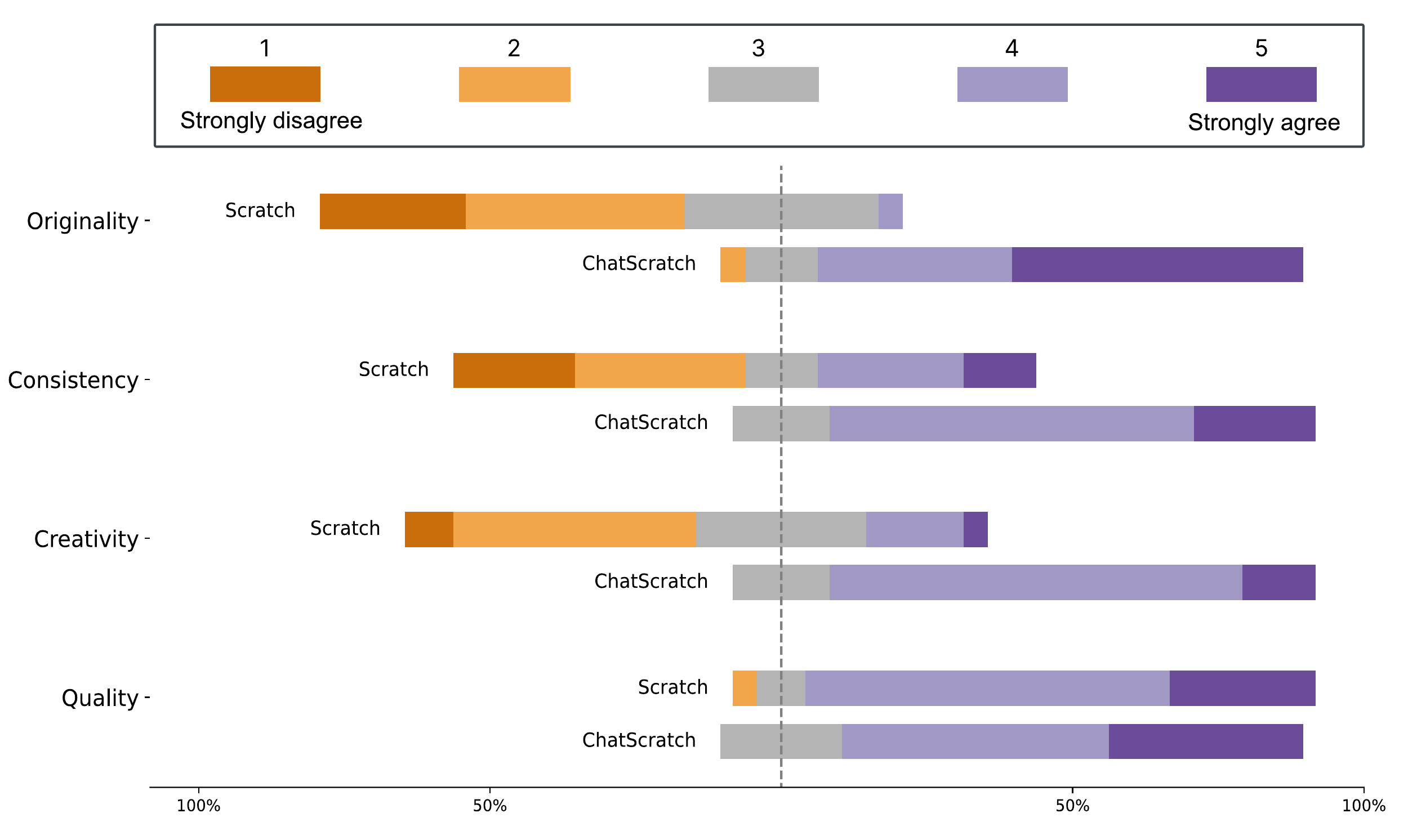}
    \vspace{-0.11in}
    \caption{\textbf{Comparison of expert ratings between Scratch and ChatScratch, based on originality, consistency, creativity, and quality of assets. Each row depicts the distribution of 24 scores for a specific aspect. The distribution's positioning towards the right indicates a higher overall performance in that aspect.}}
    \vspace{-0.11in}
    \label{fig:expert_rating}
\end{figure*}

Figure \ref{fig:expert_rating} presents the expert ratings. With a paired-samples \(t\)-test in Table \ref{tab:expert rate}, we can observe an appreciation from experts for assets children created using our ChatScratch. Significant differences were noted in originality \(t(23) = 13.06, p < 0.01^{**}\), consistency \(t(23) = 4.22, p < 0.01^{**}\), and creativity \(t(23) = 6.33, p < 0.01^{**}\) between ChatScratch and Scratch. While for quality, we observed closely matched means (4.13 for ChatScratch and 4.08 for Scratch) and standard deviations (0.74 for ChatScratch and 0.72 for Scratch) which confirms that ChatScratch-produced assets are on a quality level comparable to those from Scratch's built-in library. These results suggest that with the support of visual cues and image polish, children are able to create more freely (originality) and produce assets that align closely with their expectations (consistency), leading to an overall enhanced creative expression (creativity). For example, P5 used ChatScratch to create the story of the unlucky hen \footnote{\url{https://scratch.mit.edu/projects/887846377/}}, who turned into delicious roast chicken meat in the explosion of a planetary collision. This story contains rich and exquisite assets and interesting plot design, amused all the experts. As a counterexample, P22 in his project with Scratch used only the default cat character \footnote{\url{https://scratch.mit.edu/projects/887848481/}}. Although he employed advanced coding techniques to achieve effects like scaling, cloning, and color changing of the cat, it's apparent that the story lacked completeness and clarity in its creative concept. 

\begin{table*}[thp]
    \centering
    \footnotesize
    \renewcommand{\arraystretch}{1.2}  
    \caption{\textbf{Comparison of ChatScratch and Scratch Across Expert Ratings.}}
    \vspace{-0.11in}
    \begin{tabularx}{\textwidth}{l *{2}{>{\centering\arraybackslash}X} c *{2}{>{\centering\arraybackslash}X} c *{2}{>{\centering\arraybackslash}X}}
        \hline
        Metric & \multicolumn{2}{c}{ChatScratch} & & \multicolumn{2}{c}{Scratch} & & \multicolumn{2}{c}{Paired-t test} \\
        \cline{2-3} \cline{5-6} \cline{8-9}
        & Mean & SD & & Mean & SD & & t & p \\
        \hline
        Originality & 4.29 & 0.86 & & 2.17 & 0.87 && 13.06 & $0.000^{**}$ \\
        Consistency & 4.04 & 0.62 & & 2.79 & 1.38 && 4.22 &  $0.001^{**}$\\
        Creativity  & 3.96 & 0.55 & & 2.67 & 1.01 && 6.33 & $0.000^{**}$ \\
        Quality     & 4.13 & 0.74 & & 4.08 & 0.72 && 0.20 & 1 \\
        \hline
    \end{tabularx}
    \footnotesize
    \flushleft
    \hspace{5mm}\textbf{Notes:}
    \begin{enumerate}[label=\arabic*.]
        \item A Bonferroni correction is applied to mitigate the risk of Type I errors associated with multiple comparisons.
        \item ** denotes $p < 0.01$ and * denotes $p < 0.05$.
    \end{enumerate}
    \label{tab:expert rate}
    \Description{A horizontal bar chart that compares expert ratings between Scratch and ChatScratch across four attributes: originality, consistency, creativity, and quality. The chart uses a Likert scale from 1 (strongly disagree) to 5 (strongly agree) to represent expert consensus. The segments within each bar reflect the proportion of the 24 scores that correspond to each point on the scale. The ratings for ChatScratch generally skew towards the higher end of the scale, particularly in the areas of originality, consistency and creativity, indicating a stronger expert agreement on its higher performance in these attributes. }
    
\end{table*}

\begin{figure*}[thp]
    \centering
    \includegraphics[width=0.8\textwidth]{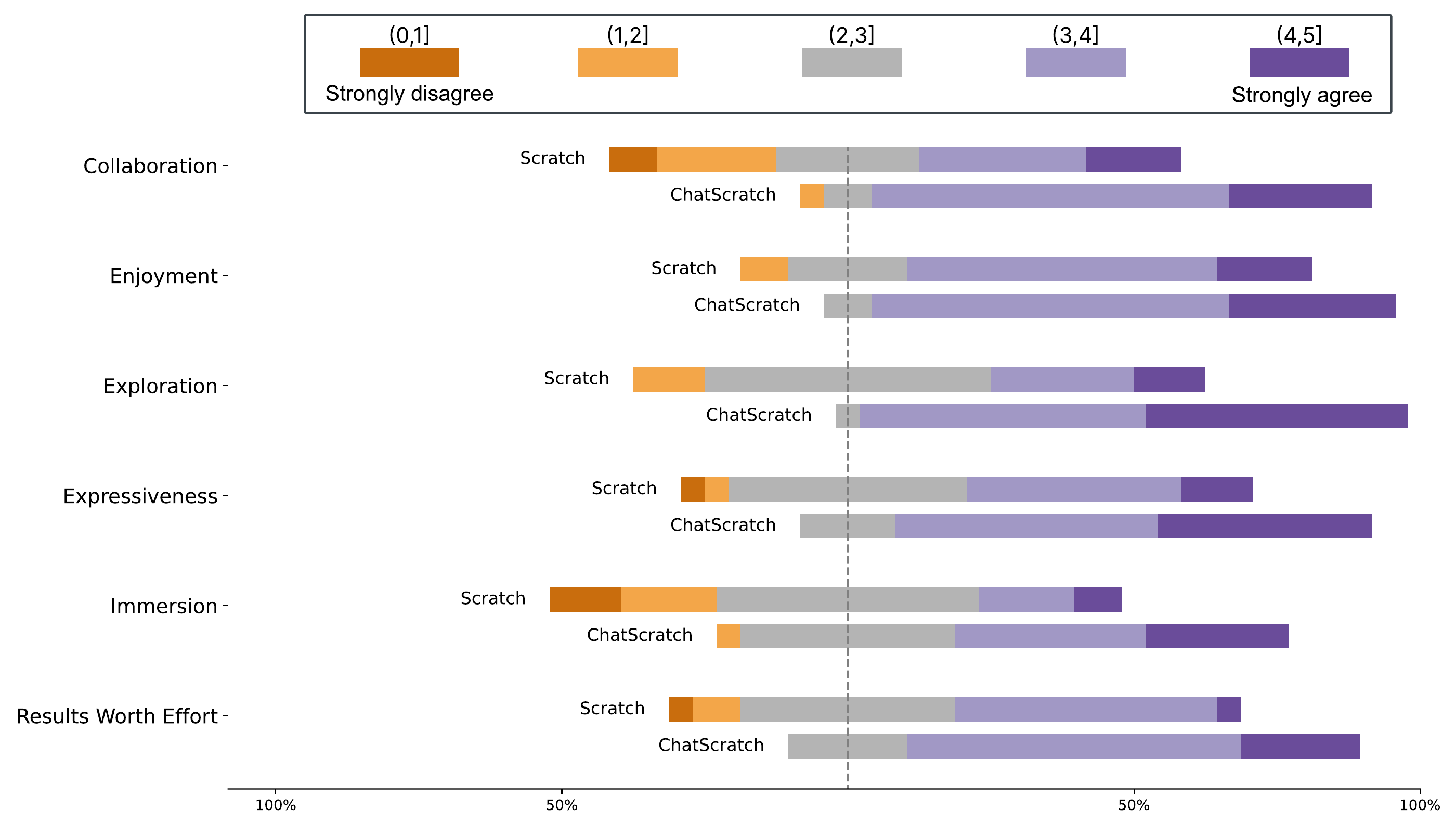}
    \vspace{-0.11in}
    \caption{\textbf{Comparison of creativity support index questionnaires between Scratch and ChatScratch, based on collaboration, enjoyment, exploration, expressiveness, immersion,  and results worth effort. Each row depicts the distribution of 24 scores for a specific aspect. The distribution's positioning towards the right indicates a higher overall performance in that aspect. }}
    \vspace{-0.11in}
    \label{fig:CSI}
    \vspace{0.11in}
\end{figure*}

We are also concerned about children's subjective perceptions of the creative support they receive. From Figure \ref{fig:CSI}, we can see that children have a positive view of both systems, and ChatScratch received higher evaluations across all six dimensions. Statistically, with a paired-samples \(t\)-test shown in Table \ref{tab:creativity support index}, we observed significant differences in the areas of collaboration \(t(23) = 3.68, p < 0.01^{**}\), exploration \(t(23) = 4.75, p < 0.01^{**}\), expressiveness \(t(23) = 3.42, p < 0.05^{*}\), immersion \(t(23) = 3.88, p < 0.01^{**}\) and results worth effort \(t(23) = 3.77, p < 0.01^{**}\). Additionally, the SD for the ChatScratch group was consistently lower than those for the corresponding Scratch group, indicating that ChatScratch provides more stable and uniform support. In summary, the data sufficiently reflects our expectations for ChatScratch. With the collaborative support provided by ChatScratch (collaboration), children's efforts in creative tasks are better represented (results worth the effort). Visual cues support children in exploring the richness and details of projects (exploration), while image polish addresses the inconvenience of original asset acquisition (expressiveness), allowing children to focus more on creation (immersion).

\begin{table*}[thp]
    \centering
    \footnotesize
    \renewcommand{\arraystretch}{1.2} 
    \caption{\textbf{Comparison of ChatScratch and Scratch Across Creativity Support Index.}}
    \vspace{-0.11in}
    \begin{tabularx}{\textwidth}{l *{2}{>{\centering\arraybackslash}X} c *{2}{>{\centering\arraybackslash}X} c *{2}{>{\centering\arraybackslash}X}}
        \hline
        Metric & \multicolumn{2}{c}{ChatScratch} & & \multicolumn{2}{c}{Scratch} & & \multicolumn{2}{c}{Paired-t test} \\
        \cline{2-3} \cline{5-6} \cline{8-9}
        & Mean & SD & & Mean & SD & & t & p \\
        \hline
        Collaboration & 3.94 & 0.74 & & 3.10 & 1.19 & & 3.68 & \(0.007^{**} \) \\
        Enjoyment & 4.00 & 0.61 & & 3.60 & 0.82 & & 2.33 & 0.175 \\
        Exploration & 4.31 & 0.59 & & 3.31 & 0.92 & & 4.75 & \(0.000^{**} \) \\
        Expressiveness & 4.08 & 0.73 & & 3.31 & 0.91 & & 3.42 & \(0.014^{*} \) \\
        Immersion & 3.63 & 0.92& & 2.77 & 1.15 & & 3.88 & \(0.004^{**}\) \\
        Results Worth Effort & 3.88 & 0.66 & & 3.17 & 0.83 & & 3.77 & \(0.006^{*}\)\\
        \hline
    \end{tabularx}
    \footnotesize
    \flushleft
    \hspace{5mm}\textbf{Notes:}
    \begin{enumerate}[label=\arabic*.]
        \item A Bonferroni correction is applied to mitigate the risk of Type I errors associated with multiple comparisons.
        \item ** denotes $p < 0.01$ and * denotes $p < 0.05$.
    \end{enumerate}
    \label{tab:creativity support index}
    \vspace{-0.11in}
\end{table*}

\subsection{RQ2: How does ChatScratch Support High-quality Code Implementation in Scratch Programming }

Another objective of our study focuses on how ChatScratch assists children in achieving higher code quality. In the first analysis, we demonstrate the results of Dr. Scratch rubrics (Table \ref{tab:dr_scratch_rubrics}), which evaluate the quality of code by assessing the level of CT displayed in children's projects. For total scores, the mean was \(13.21\) (\(SD = 2.40\)) for ChatScratch and \(8.13\) (\(SD = 2.69\)) for Scratch. This performance elevation signifies a progression from ``basic'' to nearly ``master'' level within the framework of Dr. Scratch's assessment criteria. Based on a paired-samples \(t\)-test, we observed significant differences in six dimensions: abstraction \(t(23) = 8.18, p < 0.01^{**}\), parallelism \(t(23) = 3.42, p = 0.02^{*}\), logical \(t(23) = 3.11, p = 0.03^{*}\), synchronization \(t(23) = 4.25, p < 0.01^{**}\), interactivity \(t(23) = 4.30, p < 0.01^{**}\), and data \(t(23) = 3.40, p = 0.02^{*}\), except for flow control \(t(23) = 2.50, p = 0.14\). Qualitatively, we notice a disparity in children's performance between two of the most crucial concepts in programming: conditionals (logic) and loops (flow control). Even with the support of ChatScratch which has led to notable improvements, children still lag in the area of logic. Based on our interviews, we learned that children were more comfortable with loops. They perceived the idea of using a ``repeat'' statement to carry out an action multiple times as quite intuitive. In contrast, when it came to using conditional statements such as ``if-else'', children were more hesitant. They specifically mentioned the challenge of determining the exact conditions needed in ``if-else'' statements to achieve their desired program outcomes. This reveals a greater level of complexity and abstraction involved in understanding and applying these conditionals. In addition, children also demonstrated inadequate performance in data. According to Dr. Scratch's rubrics, high scores in the data dimension require children to master the use of complex data structures like lists. This proves to be a steep learning curve for young learners, who are typically more accustomed to handling simpler types of data.

In the second analysis, two key objectives are addressed. First, we evaluate the efficacy and usability of the code produced by our code assistant, focusing on the degree to which children adopted the provided code suggestions. Second, we aim to emphasize that children's improved code quality is not solely attributed to our provided code; their own efforts are indispensable. To elucidate these points, we quantified the children's code performance during the programming activities. To illustrate the first point, we use the metric of code retention \textit{R}. With an average score of \(M = 0.74\) and \(SD = 0.15\), this metric illustrates that about three-quarters of our suggested code is retained by the children. This high retention rate indicates the usefulness of the suggested code, thereby validating the accuracy of the coding solutions generated by our code assistant. Another metric, code expansion \textit{E}, serves to highlight children's autonomous contributions to coding projects. The final projects reveal that more than half of the final code (\(M = 0.54; SD = 0.17\)) is implemented by the children themselves. This emphasizes the significance of the children's own efforts in achieving high code quality. Additionally, the data reveals that, after overcoming initial uncertainties with the foundational support provided by ChatScratch, the children were able to exhibit remarkable coding abilities. 

\subsection{RQ3: How does ChatScratch Support Children to Autonomously Create Personally Meaningful Projects}

\begin{table*}[thp]
    \centering
    \footnotesize
    \renewcommand{\arraystretch}{1.1}  
    \caption{\textbf{Comparison of ChatScratch and Scratch Across Dr. Scratch Rubrics.}}
    \vspace{-0.11in}
    \begin{tabularx}{\textwidth}{l *{2}{>{\centering\arraybackslash}X} c *{2}{>{\centering\arraybackslash}X} c *{2}{>{\centering\arraybackslash}X}}
        \hline
        Metric & \multicolumn{2}{c}{ChatScratch} & & \multicolumn{2}{c}{Scratch} & & \multicolumn{2}{c}{Paired-t test} \\
        \cline{2-3} \cline{5-6} \cline{8-9}
        & Mean & SD & & Mean & SD & & t & p \\
        \hline
        Abstraction & 2.29 & 0.46 & & 1.25 & 0.53 & & 8.18 & $0.000^{**}$ \\
        Parallelism & 2.21 & 0.78 & & 1.46 & 1.14 & & 3.42 & $0.016^{*}$ \\
        Logical & 1.29 & 1.08 & & 0.63 & 0.71 & & 3.11 & $0.034^{*}$ \\
        Synchronization & 2.13 & 0.95 & & 1.04 & 1.16 & & 4.25 & $0.002^{**}$ \\
        Flow Control & 2.17 & 0.56 & & 1.67 & 0.87 & & 2.50 & 0.139 \\
        Interactivity & 2.17 & 0.56 & & 1.46 & 0.66 & & 4.30 & $0.002^{**}$ \\
        Data & 0.95 & 0.20 & & 0.62 & 0.49 & & 3.39 & $0.018^{*}$ \\
        \hline 
        Total Score & 13.21 & 2.40 & & 8.13 & 2.69 & & 8.13 & $0.000^{**}$ \\
        \hline
    \end{tabularx}
    \footnotesize
    \flushleft
    \hspace{5mm}\textbf{Notes:}
    \begin{enumerate}[label=\arabic*.]
        \item A Bonferroni correction is applied to mitigate the risk of Type I errors associated with multiple comparisons.
        \item ** denotes $p < 0.01$ and * denotes $p < 0.05$.
    \end{enumerate}
    \label{tab:dr_scratch_rubrics}
    \Description{A horizontal bar chart comparing Scratch and ChatScratch across several creativity support index aspects: collaboration, enjoyment, exploration, expressiveness, immersion, and results worth effort. The chart uses a Likert scale from 1 (strongly disagree) to 5 (strongly agree) to represent expert consensus. The segments within each bar reflect the proportion of the 24 scores that correspond to each point on the scale. The chart shows that, in every category, ChatScratch bars are positioned predominantly towards the higher end of the scale, indicating a higher level of agreement among respondents that ChatScratch performs well in these aspects.  For ChatScratch, the results show that there are no scores in the strongly disagree range. The majority of the ratings fall above the 3-point borderline, indicating a generally positive expert consensus. }
    
\end{table*}

In this section, we discuss the findings from video recordings and interviews, which help us understand how ChatScratch better supports children in autonomously creating personally meaningful projects. 

The first topic is about how the process of iterative creation supports personally meaningful projects. When using ChatScratch, we witnessed a notable increase in iterative improvement practices among young users. This trend was evident not only during the initial stages of conceptualization and assets creation but also in the programming phase, reflecting a deeper engagement in the creative process. A case in point is P1, a 10-year-old boy, illustrates the enhanced attention to detail in children's iterative creation. He eloquently described his iterative approach in asset creation:
\begin{quote}
\textit{``I told ChatScratch I wanted to draw a Kung Fu tiger. The first time I sketched out the tiger's shape, the second time I added color, and the third time I adjusted its pose. With each modification, the generated result got closer to what I had in mind. I really enjoyed the process; it's super fun!''}
\end{quote}

Another example from P12 showcases the increased reflection and self-evaluation in the creative process. After developing several scenes, she revisited her character design before. As she put it:

\begin{quote}
\textit{``I think the color of the (protagonist's) dress isn't suitable.''} 
\end{quote}

For programming phase, an illustrative example is the experience of P13, who navigated the gap between the actual effects of given codes solution and his expectation. By identifying new requirements based on this disparity, P13 synthesized multiple coding solutions to craft a more expressive code. By incorporating visual cues, image polish, and code templates in ChatScratch, we provide children with the opportunity to receive ongoing, intermediate feedback on their creative endeavors. This allows for a gradual refinement process, ensuring that their projects progressively evolve to match their vision. Statistically, 16 out of 24 participants mentioned the contribution of visual cues to personalization, 21 highlighted the impact of image polish, and 15 acknowledged the role of code template. On average, the children used visual cues 3.12 times (SD = 1.74), image polish 7.41 times (SD = 2.33), and coding assistance 4.78 times (SD = 1.78). 

Second, we found that support personal preferences also contribute to the success of personally meaningful projects. Although our experiment was structured around two predefined themes, it became evident that children's individual preferences and passions often transcended these set boundaries, allowing unique interpretations on preset theme. For instance, P4, an 8-years-old girl, used ChatScratch for an animal story and Scratch for a personal experience story, stated:

\begin{quote}
\textit{`` I like pandas and snow leopards, so I made a story about them using ChatScratch. When I was using Scratch, I documented the experience of my mom taking photos of me. I liked the red dress I wore that day, but unfortunately, Scratch didn't have it.''}
\end{quote}

P4's narrative illustrates how children's choices and likes distinctly shape their creative outputs. Whether it was the animals she adores or the memory of a special dress, her personal inclinations were clearly reflected in her projects, thereby making the work uniquely her own. Building on the theme of personal preferences shaping children's projects, another observation from our study further reinforces this idea. We noticed a fascinating trend among three children (P2, P7, P14) who all chose to develop their animal stories based on the classic fable of \textit{The Tortoise and the Hare}. However, each child introduced a distinct twist, infusing their unique interests into the narrative. P2, for instance, set their story in a vibrant urban environment, adding a contemporary flair. P7, drawing inspiration from a wildlife documentary, replaced the classic characters with a Tibetan fox and a plateau pika, showcasing a blend of traditional storytelling with modern ecological awareness. Meanwhile, P11 elevated the tale by adding an interactive dimension, turning the story into a game where characters could be maneuvered using a mouse and keyboard. Their stories, while rooted in a same origin, diverged in ways that mirrored their unique perspectives and preferences, showcasing the powerful impact of personal interests in shaping creative expressions.

\section{Discussion}
In this section, we further discuss our research questions, providing explicit answers based on the results of our experiments. Firstly, we demonstrate how ChatScratch effectively leverages the advantages of generative artificial intelligence to overcome the challenges identified in our formative study, thereby providing substantial support for creativity (RQ1) and programming skill (RQ2) in children's visual programming activities. Secondly, we discuss the importance of providing support for both creativity and programming during the creation of personally meaningful projects (RQ3), and how such an approach is significant for enhancing autonomous learning. Finally, we explore some limitations revealed by ChatScratch and discuss directions for future work.

\subsection{The Benefit of Generative Artificial Intelligence in ChatScratch}

In our ChatScratch, the generative artificial intelligence (GAI) plays a crucial role in assisting children with project planning, supporting their creative endeavors, and providing guidance in programming. In the preparation phase, ChatScratch distinguishes itself by attentively addressing creative barriers encountered by children during their creative process. It aids not only in refining existing material but also in generating new content that is closely linked to what has already been created. To achieve this, ChatScratch must comprehend children's verbal story expressions, expand on them, and then transform these into prompts suitable for a visual generation module. This process results in outputs that are more concrete and easily understandable visual representations of story concepts, facilitating children's comprehension. This capability is precisely where GAI excels—it possesses the ability to understand, reason, and think creatively. These skills are crucial for guiding and enriching the creative process in children, especially in an autonomous learning environment where outside resources may be limited. Under the assistance of ChatScratch, children participating in the experiment created an average of 1.75 more assets. Furthermore, feedback from the CSI questionnaire indicated that they found the entire creative process to be smoother and more effective.

In the assets creation phase, our primary goal is to assist children in collaboratively creating high-quality assets that align with their aspirations. Based on our formative study, ChatScratch has achieved the three objectives we set. First, as part of the programming learning experience, children, as the central figures in the creative activity, spearhead the progression of the entire process. They freely engage in drawing and describing assets through multiple iterations and feedback, steering their own creative exploration. Second, the contents  generated by our system must be non-toxic and safe, suitable for children. Third, the created assets need to meet children's expectations, fulfilling their vision for specific themes or images, thereby achieving their sense of accomplishment and engagement. 

With the enhancement of GAI, ChatScratch also fulfills the requirement for high-quality code implementation. In achieving this objective, the key necessity, as identified in our formative study, is to provide a simple and easy starting point, thereby reducing the learning curve for the task. The incorporation of structured Scratch card knowledge has enhanced ChatScratch's ability to comprehend the Scratch language, enabling it to excellently perform the mapping from task requirements to the implementation of code blocks. Additionally, a feature of ChatScratch that greatly intrigued and thrilled the children was its dual functionality in code implementation: it goes beyond merely suggesting how to piece together code blocks, to actually creating and inserting a fully executable script directly into the coding area, allowing for immediate trial and interaction. However, as we did not want AI to replace the children's efforts in the learning process, the code implementation provided by ChatScratch was intentionally designed as merely a skeleton. This approach encourages children to modify, populate, and enrich it, ultimately leading them to craft their own unique implementation. The code expansion metric in our evaluation demonstrates that the children excelled in this aspect. In the final project code, 57\% of the code blocks were created by the children themselves, effectively building upon the foundational structure we provided.

\subsection{Cultivating Creativity and Programming Skills in Children through ChatScratch}
Rooted in Scratch's core philosophy of blending creativity with programming, it's vital to address both aspects in supporting children's Scratch learning. This concept steered ChatScratch's development, emphasizing enhancement of creative tasks including project planning and assets creation. 

Creativity serves as a key enhancer in the programming process. Programming with Scratch resembles constructing with LEGO bricks. Developing coding proficiency is akin to mastering the art of arranging blocks to create structures with specific functions  – a vital skill indeed. However, the more significant aspect that determines the result is what you intend to build. In our experiments, we observed a compelling connection between the children's engagement in their storytelling and their coding creativity. The project planning feature in Scratch provides a structured method for story development, prompting children to think logically and generate diverse ideas for expression through coding. Furthermore, ChatScratch's asset creation feature, which provides customizable characters and scenes, enables children to develop a strong attachment to their characters. Using ChatScratch, children's projects become more imaginative, leveraging a range of coding strategies to animate their stories, such as character movements, dialogues, and visual appearance changes. Moreover, a robust creative idea instills resilience. Driven by the aspiration to realize their vision as imagined, children are less likely to give up. They persevere through intricate coding challenges, learning from errors and continuously refining their projects.

Combining support for both creativity and programming leads to higher completion rates and more engaging personally meaningful projects. In such cases, children are more immersed, willing to share, and feel a greater sense of participation, reflecting their investment and enthusiasm. This heightened engagement and satisfaction are crucial in sustaining interest in ChatScratch, thereby encouraging its consistent use and practice. Such continual interaction is vital for developing and mastering visual programming skills, particularly in the context of autonomous learning.

\subsection{Limitations and Future Work}
In this part, we present two limitations of our ChatScratch. First, drawing from prior research, ChatScratch facilitates voice-based interactions for children, thereby addressing potential literacy problems they may encounter. However, we found in our experiments that when issues arise during interaction, the absence of visual input feedback might lead children to incorrectly assume that these problems stem from their own mistakes, rather than from the system's limitations or its way of processing inputs. This might not have been a significant problem in previous system, where the tasks handled were generally more straightforward and followed a more predictable pattern. In systems incorporating LLMs like ChatScratch, the situation is indeed different. These systems are expected to handle more complex tasks, and the system itself is more flexible and sensitive in processing inputs. To solve this problem, potential solutions in the future include rephrasing or refining children's inputs, or asking additional clarifying questions. 

In ChatScratch, we also introduced a structured project plan framework, segmenting children's projects into three main acts, with an emphasis on key elements including characters, scenes, and events. Overall, such a design is beneficial for children as it reduces their cognitive load during the conceptualization phase. However, we must acknowledge that this design inherently limits the ChatScratch system's generalizability. Specifically, the three-act structure of ``Setup - Development - Climax'' or ``Introduction - Conflict - Resolution'' is more suited to stories with strong conflicts and twists, such as adventure tales that are often favored by children. When children seek to craft linear, descriptive stories, like narrating a typical day in their lives, the structured three-act design might not be as suitable. In addition, some participants mentioned in interviews their desire to use ChatScratch for creating other types of works, such as music and games. These areas, however, are currently not covered in our system.

Beyond addressing the existing limitations, we have the following expectations for the future development of ChatScratch. On one hand, we recognize the necessity of providing more comprehensive guidance in the key area of coding instruction. In addition to offering basic implementation solutions currently included in the system, guiding children through task decomposition to progressively develop their coding skills is another avenue to explore. A potential approach includes utilizing interactive mind maps to assist children in constructing complete code implementation paths. On the other hand, the current system lacks mechanisms for assessing and providing feedback on the autonomous learning process. How to code and quantify information about children's learning processes, in order to understand whether they truly benefit from using the system, is an aspect that needs consideration.
\section{Conclusion}

In this paper, we presented ChatScratch, an AI-augmented system designed for autonomous visual programming learning, targeting primary school children aged 6-12. We conducted a formative investigation involving six experienced educators and participated in offline programming classes to understand the challenges faced by children and to identify implications for design. Our system is powered by generative models that can generate visual cues, polish sketch drawings, and provide code assistance. In our within-subject study involving 24 children, we found that interactive structured storyboards and visual cues significantly enhance the richness and detail of children's projects. Digital drawing with image polish techniques supports the digital asset creation process. Moreover, a Scratch-specialized large language model assists children in their coding journey and improves their computational thinking skills. We discussed insights from our findings that pave the way for children to create personally meaningful projects with generative models. These insights also support the provision of real-time, hands-on, and autonomous programming learning experiences.
\begin{acks}
We express our sincere gratitude to all the reviewers for their invaluable guidance and suggestions; Gang Dai, Wen Meng and Xuenan Jiang from AC Source Programming for their expert educational perspectives and assistance with the experiment; all the user study participants for their time and contribution; Yu Cai for providing valuable insights on data analysis. This research was funded by National Natural Science Foundation of China (Grant No. 62207023), and The Ng Teng Fong Charitable Foundation in the form of ZJU-SUTD IDEA (Grant No. 188170-11102). 
\end{acks}


\end{document}